\documentclass[aps,prd,
superscriptaddress,
showkeys,
showpacs,
preprint]{revtex4-1}

\usepackage[english]{babel}
\usepackage{amsmath,bm}
\usepackage{amssymb}
\usepackage{enumitem}
\usepackage{textcomp}
\usepackage{graphicx}
\usepackage{dsfont}
\usepackage{color}
\usepackage{hyperref}

\usepackage{mathrsfs}
\usepackage{array}

\definecolor{purple}{rgb}{0.8,0,0.6}

\graphicspath{{Figures/}}

\begin{document}
\title{Electrical conductivity of hot Abelian plasma \\
	with scalar charge carriers}

\author{O.O.~Sobol}
\affiliation{Institute of Physics, Laboratory for Particle Physics and Cosmology (LPPC), \'{E}cole Polytechnique F\'{e}d\'{e}rale de Lausanne (EPFL), CH-1015 Lausanne, Switzerland}
\affiliation{Physics Faculty, Taras Shevchenko National University of Kyiv, 64/13, Volodymyrska Street, 01601 Kyiv, Ukraine}

\date{\today}

\begin{abstract}
	We study the electrical conductivity of hot Abelian plasma containing scalar charge carriers in the leading logarithmic order in coupling constant $\alpha$ using the Boltzmann kinetic equation. The leading contribution to the collision integral is due to the M{\o}ller and Bhabha scattering of scalar particles with a singular cross section in the region of small momentum transfer. Regularizing this singularity by taking into account the hard thermal loop corrections to the propagators of intermediate particles, we derive the second order differential equation which determines the kinetic function. We solve this equation numerically and also use a variational approach in order to find a simple analytical formula for the conductivity. It has the standard parametric dependence on the coupling constant $\sigma\approx 2.38\, T/(\alpha \log\alpha^{-1})$ with the prefactor taking a somewhat lower value compared to the fermionic case. Finally, we consider the general case of hot Abelian plasma with an arbitrary number of scalar and fermionic particle species and derive the simple analytical formula for its conductivity.
\end{abstract}


\maketitle

\section{Introduction}
\label{sec-intro}

Transport coefficients are very important characteristics of any medium providing information about its response to external perturbations. In particular, electrical conductivity is a basic property of any material describing the electric charge transfer, if an external electric field is applied to the system. Whereas in strongly coupled systems it is usually impossible to predict theoretically this coefficient and it is determined experimentally, in a weakly coupled quantum field theory it can be calculated from the first principles. One of the well-known examples of the latter system is the weakly-coupled plasma which exists in the Universe on different stages of its evolution. Knowing the conductivity of such a plasma is important for different cosmological applications, such as describing the evolution of primordial magnetic fields \cite{Turner:1988,Kronberg:1994,Joyce:1997,Grasso:2001,Durrer:2013,Subramanian:2016}, the processes of lepto- and baryogenesis \cite{Cohen:1993,Rubakov:1996,Giovannini:1998,Kamada:2016a,Kamada:2016b}, the evolution of chiral asymmetry \cite{Boyarsky:2012,Figueroa:2018a,Figueroa:2018b,Figueroa:2019} etc.

There are two ways to calculate the transport coefficients from first principles. The first one is a microscopic approach based on the linear response theory, which allows one to express the transport coefficients in terms of retarded correlation functions of the conserved currents in the low momentum and frequency limit (the so-called Kubo relations). These correlators have to be calculated in finite temperature quantum field theory and the leading order result is obtained by the resummation of an infinite number of ladder diagrams. This approach was applied to the calculation of the transport coefficients in scalar field theory \cite{Hosoya:1984,Jeon:1995,Jeon:1996}, in gauge theories \cite{Basagoiti:2002,Gagnon:2006} as well as in the effective models of quantum chromodynamics (QCD) such as the Nambu-Jona-Lasinio model \cite{Harutyunyan:2017,Ghosh:2019}, the lattice QCD \cite{Aarts:2015}, the Polyakov-Quark-Meson model \cite{Singha:2019} etc. Although it is the most general approach, applicable to any given field theory, it requires inventing ingenious resummation schemes, especially for the gauge theories, and therefore it is not very convenient.

On the other hand, there is also another approach based on kinetic theory. The Boltzmann equation appeared in classical physics and operates with pointlike particles rather than quantum fields. Nevertheless, the behavior of the hard modes in quantum field theory in the weak-coupling limit admits a description in terms of quasiparticles (throughout the paper we will call the momenta $k\gtrsim T$ as \textit{hard}, and those $k\ll T$ as \textit{soft} ones). Indeed, starting from the Schwinger-Dyson (SD) equations for the real-time Green's functions at finite temperature and assuming a small departure from equilibrium it is possible to show in the weakly coupled theory, that the hard modes can be regarded as massless quasiparticles interacting through a screened potential. Then, performing the gradient expansion in SD equations and introducing the Wigner functions which have the meaning of the phase space distribution, it is easy to show that they satisfy the Boltzmann equation with the collision integral in a canonical form including gain and loss terms \cite{Calzetta:1988,Blaizot:1993,Blaizot:1994,Blaizot:1999,Calzetta:2000,Blaizot:2002}. The scattering amplitudes in the collision integral have to be calculated using the Feynman rules of the corresponding quantum field theory. The kinetic equation describes the deviation from equilibrium of the hard particles distribution function on scales large compared to thermal de Broglie wavelength $\lambda_{dB}\sim 1/T$. 

The kinetic approach is much simpler than the diagrammatic one, but unfortunately, it allows one to calculate the transport coefficients only in the leading order in the coupling constant. To compute the higher order corrections, one has to take into account the modification of the dispersion relations of the hard particles coming from the self-energy corrections to their propagators. Usually the leading order result is satisfactory for most of the practical issues and the kinetic approach is widely used to calculate the transport coefficients. In particular, the electrical conductivity of the ultrarelativistic plasma was computed in the framework of kinetic theory in Refs.~\cite{Gavin:1985,Czyz:1986,Heiselberg:1992,Oertzen:1992,Ahonen:1996,Baym:1997,Ahonen:1998,Arnold:2000,Arnold:2003,Arnold:2003-col,Puglisi:2014}.

First attempts to extract the conductivity from kinetic approach were made using the simplest $\tau$-approximation for the collision integral \cite{Gavin:1985,Czyz:1986, Heiselberg:1992} or performing the Chapman-Enskog expansion \cite{Oertzen:1992}. Later, the Boltzmann equation was numerically solved using the Monte Carlo simulations of the collision integral, and the conductivity of the early Universe was determined in Refs.~\cite{Ahonen:1996,Ahonen:1998}. The leading-log result for the conductivity of Abelian plasma was derived in the constant flow velocity approximation in Ref.~\cite{Baym:1997}, however not including the Compton scattering and pair annihilation processes which are equally important as the M{\o}ller and Bhabha scattering in the fermionic case. This inconsistency was removed in Ref.~\cite{Arnold:2000} where the leading-log expressions for the conductivity, viscosity, and flavor diffusivity of the hot fermion plasma were derived in a variational approach. Finally, in the companion paper \cite{Arnold:2003}, the corresponding results in the next-to-leading-log approximation as well as in the full leading order were presented.

Electrical conductivity in the early Universe in the Standard Model framework was discussed in Refs.~\cite{Ahonen:1996,Baym:1997,Ahonen:1998,Arnold:2000,Arnold:2003}. A nonzero charge of a particle with respect to $U(1)$ gauge interaction makes it an effective scatterer for other charge carriers. However, the electric current is determined only by those particles for which the $U(1)$ interaction is the strongest one among all interactions they can take part. Since the scalar Higgs particle interacts in the symmetric phase not only with $U(1)_{Y}$ hypercharge gauge field but also (more intensively) with $SU(2)_{I}$ gauge bosons, the departure from equilibrium caused by the hyperelectric field is washed out more quickly by means of $SU(2)$ interactions and the contribution to the hyperelectric current is negligible. On the other hand, in the broken symmetry phase, the Higgs boson is neutral and plays no role in the electrical conductivity. 

Charged scalar particles also exist in hadron gas for the temperatures below the deconfinement phase transition in QCD. In Refs.~\cite{Fraile:2006,Nicola:2007,Lee:2014,Greif:2016,Ghosh:2017,Atchison:2017,Kadam:2018,Ghosh:2018}, the electric conductivity of such a plasma was calculated by the variety of methods, including Kubo relations, kinetic approach, chiral perturbation theory etc. However, the lightest scalars, pions, have the mass $m_{\pi}=138\,{\rm MeV}$ which is of the same order as the phase transition temperature $T_{c}\simeq 150\,{\rm MeV}$. This means that in the hadronic phase one cannot treat these particles as ultrarelativistic ones and have to take into account their masses. Moreover, the mean-free path of charged hadrons is mainly determined by their strong interaction, and the simplified description in the framework of scalar electrodynamics is insufficient.

Therefore, the kinetic equation for ultrarelativistic scalar charged particles was not studied previously in the literature and their contribution to the conductivity was not calculated. This could be interesting, however, in some extensions of the Standard Model containing extra charged scalars. Also it may be useful to compare the results of numerical lattice simulations including scalars with theoretical predictions. In particular, in Refs.~\cite{Figueroa:2018a,Figueroa:2018b,Figueroa:2019} the fermion chirality nonconservation in Abelian gauge theory with a charged scalar field was studied. The theoretical prediction for the chirality breaking rate depends on the electrical conductivity in the system, and it has to be compared with the numerical result measured from lattice simulations. Finally, it is important to study the role of the scalar particles in order to understand the whole picture of the transport phenomena in hot Abelian plasmas.  

This paper is organized as follows.  We consider the general form of the Boltzmann kinetic equation with the collision integral resulting from $2\leftrightarrow 2$ scatterings in Sec.~\ref{sec-Boltzmann}. In Sec.\ref{sec-conductivity-sQED}, the electrical conductivity of the hot scalar QED plasma with one type of charge carriers is calculated using the exact solution of the Boltzmann equation as well as by variational method. In Sec.~\ref{sec-general} we derive the general formula for the conductivity of a multicomponent Abelian plasma with an arbitrary number of charged scalar and fermionic particles. The summary of the obtained results is given in Sec.~\ref{sec-conclusion}. In Appendix~\ref{sec-app-details} we provide some details of the calculation of the collision integral in the leading-log order.

\section{Boltzmann equation}
\label{sec-Boltzmann}

As we mentioned in the Introduction, in order to calculate any transport coefficient at leading order in the coupling constant, it is sufficient to use the Boltzmann kinetic equation which describes the dynamics of the phase space distribution of hard particles in the plasma. We will follow the standard kinetic approach previously used in calculations of the conductivity in the fermionic case in Refs.~\cite{Ahonen:1996,Ahonen:1998,Baym:1997,Arnold:2000}. Let us consider the Abelian plasma which consists of several constituents, which we will mark by the small Latin index $a,\, b,\, c$, etc. In particular, in scalar electrodynamics, there are three of them: scalar particles, their antiparticles, and photons. Each constituent can be characterized by a phase space distribution function $f^{a}(X,\mathbf{k})$, where $X=(t,\ \mathbf{x})$ marks a space-time point and $\mathbf{k}$ is the particle's momentum. Then, the Boltzmann equation reads:
\begin{equation}
\label{Boltzmann-eq}
\left(\partial_{t}+\mathbf{v}\cdot\boldsymbol{\nabla}_{x}+\mathbf{F}\cdot\boldsymbol{\nabla}_{k}\right)f^{a}(X,\mathbf{k})=-\mathcal{C}^{a}[f],
\end{equation}
where $\mathbf{v}=\mathbf{k}/|\mathbf{k}|$ is a particle velocity, and $\mathbf{F}$ is an external force, e.g. the electric force in the case we are studying the conductivity.
In order to calculate the transport coefficients in the leading-log approximation it is sufficient to consider only $2\leftrightarrow 2$ processes, e.g. the M{\o}ller, Bhabha or Compton scattering, pair annihilation and so on. Indeed, as it was discussed in the fermionic case in Refs.~\cite{Baym:1997,Arnold:2000}, the additional logarithmic dependence on the coupling constant originates from the IR logarithmic divergence in the matrix element corresponding to these scattering processes. The processes $2\leftrightarrow n$ with $n>2$, in general give a result parametrically of higher order in the coupling constant. Nevertheless, it was shown in Refs.~\cite{Arnold:2003,Arnold:2003-col} that effective $1\leftrightarrow 2$ processes of nearly collinear bremsstrahlung and pair production or annihilation in the presence of soft gauge field excitations (in fact, these are the $N+1\leftrightarrow N+2$ processes including $N$ soft intermediate photons) may also contribute in the same order as $2\leftrightarrow 2$ processes. However, they do not have the additional logarithmic enhancement and should be taken into account only in the full leading order calculation, which lies beyond the scope of our paper.

Therefore, the collision integral for the particle of type $a$ has the form:
\begin{multline}
\label{collision-term-general}
\mathcal{C}^{a}[f]=\sum_{\{bcd\}}\int\frac{d^{3}\mathbf{k}'}{(2\pi)^{3}}\frac{d^{3}\mathbf{p}}{(2\pi)^{3}}\frac{d^{3}\mathbf{p}'}{(2\pi)^{3}}\frac{\left|\mathcal{M}^{ab}_{cd}(KP\to K'P')\right|^{2}}{16\epsilon_{k}\epsilon_{k'}\epsilon_{p}\epsilon_{p'}}(2\pi)^{4}\delta^{(4)}(K+P-K'-P')\\
\times\left[f^{a}(\mathbf{k})f^{b}(\mathbf{p})(1\pm f^{c}(\mathbf{k}'))(1\pm f^{d}(\mathbf{p}'))-(1\pm f^{a}(\mathbf{k}))(1\pm f^{b}(\mathbf{p}))f^{c}(\mathbf{k}')f^{d}(\mathbf{p}')\right],
\end{multline}
where $K=(k^{0},\,\mathbf{k})$ is the 4-momentum, with $k^{0}=\epsilon_{k}=|\mathbf{k}|$ (although, in general, the particles have a nonzero mass, in the hot plasma with temperature $T\gg m$ the hard particles with $k\gtrsim T$ can be treated effectively as massless). The delta-function takes into account the energy-momentum conservation in scattering, and the $\pm$ sign corresponds to bosonic or fermionic particle statistics, respectively. The sum over the particle species means the sum over all possible collision processes for particle $a$. $|\mathcal{M}|^{2}$ is the corresponding matrix element which takes into account the symmetry factors in the case of identical particles in the final state and number of spin or polarization degrees of freedom. Finally, the two terms in the square brackets take into account the fact that the particle of the type $a$ with momentum $\mathbf{k}$ can disappear in the scattering event $a+b\to c+d$ (the loss term) or it can appear in the inverse process $c+d\to a+b$ (the gain term). It has to be mentioned that all distribution functions appearing in the collision term are taken at the same space-time point $X$ and this dependence was omitted in Eq.~(\ref{collision-term-general}) for the sake of simplicity.

In order to compute the transport coefficients, we have to consider a slight departure from equilibrium and linearize the collision integral with respect to this deviation. Then, it is convenient to use the following decomposition:
\begin{equation}
\label{definition-W}
f^{a}(X,\mathbf{k})=f^{a}_{\rm eq}(\epsilon_{k})+\frac{\partial f^{a}_{\rm eq}}{\partial \epsilon_{k}} W^{a}(X,\mathbf{k})=f^{a}_{\rm eq}(\epsilon_{k})-\frac{1}{T}f^{a}_{\rm eq}(\epsilon_{k})(1\pm f^{a}_{\rm eq}(\epsilon_{k})) W^{a}(X,\mathbf{k}),
\end{equation}
where $f^{a}_{\rm eq}(x)=n_{B,F}(x)=[\exp(x/T)\mp 1]^{-1}$ is the Bose-Einstein or Fermi-Dirac equilibrium distribution function depending on the statistics of the particles of the type $a$. To first order in $W$, we can equivalently rewrite $f^{a}(X,\mathbf{k})\approx f^{a}_{\rm eq}(\epsilon_{k}+W^{a})$, which explains the physical meaning of the function $W$ as the local modification of the particle's dispersion relation under the impact of external force. 

We now put decomposition (\ref{definition-W}) into expression (\ref{collision-term-general}) and keep only the terms up to the first order in $W$. First of all, it is worth noting that the collision integral calculated with the equilibrium distribution functions (i.e. the contribution from the zeroth order in $W$) identically vanishes. This can be seen directly from the following identity:
\begin{equation}
\frac{f^{a}_{\rm eq}(\epsilon_{k})f^{b}_{\rm eq}(\epsilon_{p})(1\pm f^{c}_{\rm eq}(\epsilon_{k'}))(1\pm f^{d}_{\rm eq}(\epsilon_{p'}))}{(1\pm f^{a}_{\rm eq}(\epsilon_{k}))(1\pm f^{b}_{\rm eq}(\epsilon_{p}))f^{c}_{\rm eq}(\epsilon_{k'})f^{d}_{\rm eq}(\epsilon_{p'})}=\exp(-\epsilon_{k}-\epsilon_{p}+\epsilon_{k'}+\epsilon_{p'})=1,
\end{equation}
which is a manifestation of the detailed balance principle in equilibrium. Therefore, the collision integral appears to be a linear functional of the $W$-functions:
\begin{multline}
\label{collision-term-linearized}
\mathcal{C}^{a}[W]=-\frac{1}{T}\sum_{\{bcd\}}\int\frac{d^{3}\mathbf{k}'}{(2\pi)^{3}}\frac{d^{3}\mathbf{p}}{(2\pi)^{3}}\frac{d^{3}\mathbf{p}'}{(2\pi)^{3}}\frac{\left|\mathcal{M}^{ab}_{cd}(KP\to K'P')\right|^{2}}{16\epsilon_{k}\epsilon_{k'}\epsilon_{p}\epsilon_{p'}}(2\pi)^{4}\delta^{(4)}(K+P-K'-P')\\
\times f^{a}_{\rm eq}(\epsilon_{k})f^{b}_{\rm eq}(\epsilon_{p})(1\pm f^{c}_{\rm eq}(\epsilon_{k'}))(1\pm f^{d}_{\rm eq}(\epsilon_{p'}))\left[W^{a}(X,\mathbf{k})+W^{b}(X,\mathbf{p})-W^{c}(X,\mathbf{k}')-W^{d}(X,\mathbf{p}')\right].
\end{multline}

Further, we rewrite the left-hand side of the Boltzmann equation (\ref{Boltzmann-eq}) in terms of $W$ assuming the external force which pushes the particles from equilibrium to be of the same order as $W$. Then, we obtain the following:
\begin{equation}
-\frac{1}{T}f^{a}_{\rm eq}(\epsilon_{k}) (1\pm f^{a}_{\rm eq}(\epsilon_{k}))\left[v\cdot\partial_{X}W^{a}(X,\mathbf{k})+\mathbf{v}\cdot\mathbf{F}\right],
\end{equation}
where $v=(1,\,\mathbf{v})$ is the particle's 4-velocity.

In order to calculate the electrical conductivity of the plasma, we apply a constant electric field $\mathbf{E}$ to this system. Then, it is natural to assume that the kinetic function $W^{a}$ which describes the deviation of the particles distribution function from equilibrium does not depend on time and coordinates, i.e. we consider the steady state currents in the plasma. Furthermore, the charge conjugation symmetry requires for the departure from equilibrium for particles and antiparticles to have equal absolute values and opposite signs. In particular, this gives $W^{\bar{a}}=-W^{a}$ and $W^{\gamma}=0$. 

The conductivity can be extracted in a standard way by computing the electric current:
\begin{multline}
\label{electric-current}
\mathbf{j}=e\sum_{a} q_{a}g_{a}\int\frac{d^{3}\mathbf{k}}{(2\pi)^{3}}\mathbf{v} \left[f^{a}(\mathbf{k})-f^{\bar{a}}(\mathbf{k})\right]\\
=-\frac{2e}{T}\sum_{a} q_{a}g_{a}\int\frac{d^{3}\mathbf{k}}{(2\pi)^{3}}f^{a}_{\rm eq}(\epsilon_{k})(1\pm f^{a}_{\rm eq}(\epsilon_{k}))\mathbf{v} W^{a}(\mathbf{k})=\sigma \mathbf{E},
\end{multline}
where $q_{a}$ and $g_{a}$ are the charge and number of spin degrees of freedom for the particles of the type $a$, respectively, and the sum is taken over the types of charged particles not including their antiparticles separately. In the derivation we used decomposition (\ref{definition-W}) and the charge-conjugation symmetry. Knowing the kinetic function $W^{a}$ for each type of particle in the plasma, we can calculate the electric current and identify the conductivity.

\section{Electrical conductivity of scalar QED plasma}
\label{sec-conductivity-sQED}

In this section we will calculate the conductivity in scalar quantum electrodynamics (QED). We consider the self-interacting complex scalar field $\phi$ coupled to the $U(1)$ gauge field $A_{\mu}$, i.e. the electromagnetic field. The corresponding Lagrangian density has the form:
\begin{equation}
\label{lagrangian-sQED}
\mathcal{L}=-\frac{1}{4}F_{\mu\nu}F^{\mu\nu}+\left(D_{\mu}\phi\right)^{\dagger}\left(D^{\mu}\phi\right)-m^{2}\phi^{\dagger}\phi-\lambda (\phi^{\dagger}\phi)^{2},
\end{equation}
where $F_{\mu\nu}=\partial_{\mu}A_{\nu}-\partial_{\nu}A_{\mu}$ is the electromagnetic field tensor, $D_{\mu}=\partial_{\mu}-ie A_{\mu}$ is a covariant derivative, $e$ is the scalar particle's charge, and $\lambda$ is a dimensionless self-coupling constant of the scalar field. This theory allows for the three types of interaction vertices: $\phi\phi\bar{\phi}\bar{\phi}$ with coupling $\lambda$, $\phi\bar{\phi}\gamma\gamma$ with coupling $e^{2}$, and $\phi\bar{\phi}\gamma$ with coupling $e$, where by $\phi$, $\bar{\phi}$, and $\gamma$ we denote the scalar particle, antiparticle, and photon, correspondingly. In what follows we assume $\lambda\sim e^{2}$ because, in this case, all tree-level diagrams contribute to the matrix elements at the same order in $e$.

\subsection{Collision term}
\label{subsec-collision}

Let us consider all possible $2\leftrightarrow 2$ scattering processes in which the scalar particle $\phi$ takes part. They are listed in Table~\ref{tab-sqed} together with the corresponding matrix elements which are expressed in terms of the Mandelstam variables $s=(P+K)^{2}$, $t=(K-K')^{2}$, and $u=(K-P')^{2}$. 

\begin{table}[!h]
	\begin{tabular}{m{.5cm} m{4cm} m{5cm} m{5cm}}
		\hline\hline
		\begin{center}\textnumero \end{center} & \begin{center}Process \end{center} & \begin{center} Diagrams \end{center}
		& \begin{center}Matrix element $\overline{|\mathcal{M}|^{2}}$\end{center} \\
		\hline
		% 1 row
		\begin{center}1\end{center}& \begin{center}M{\o}ller scattering\ \ \  $\phi\phi \longrightarrow \phi\phi$ \end{center} &
		\vspace*{0.2cm}\includegraphics[height=1.4cm]{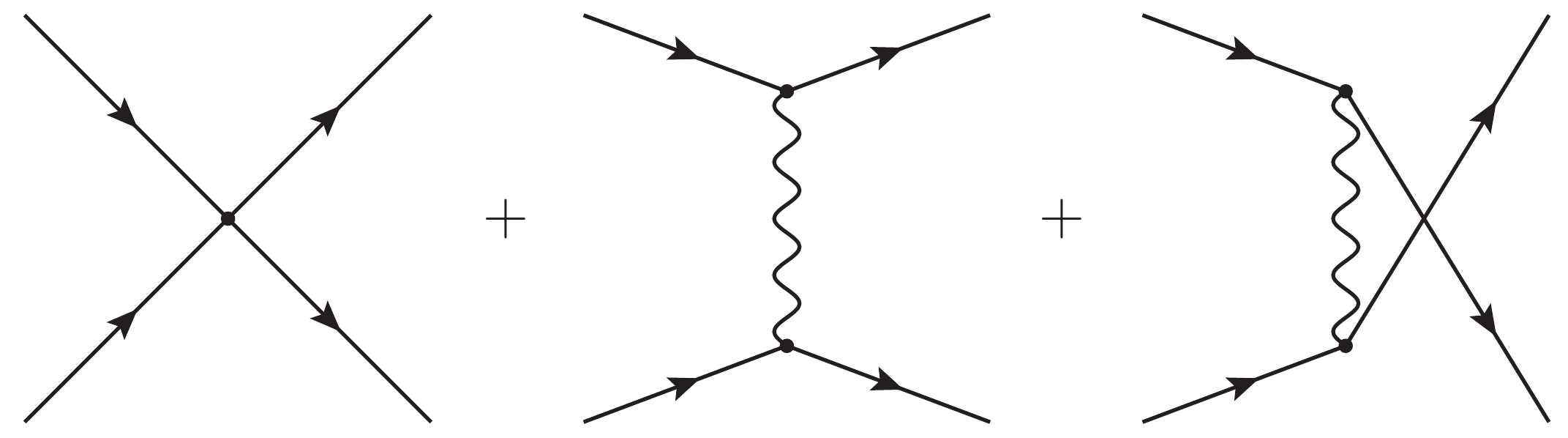}&
		\begin{center}$ \Big(4\lambda+e^{2}\dfrac{s^2+t^2+u^{2}}{tu}\Big)^{2}$\end{center}
		\\ 
		% 2 row
		\begin{center}2\end{center}& \begin{center} Bhabha scattering $\phi \bar{\phi} \longrightarrow  \phi \bar{\phi} $ \end{center}&
		\includegraphics[height=1.4cm]{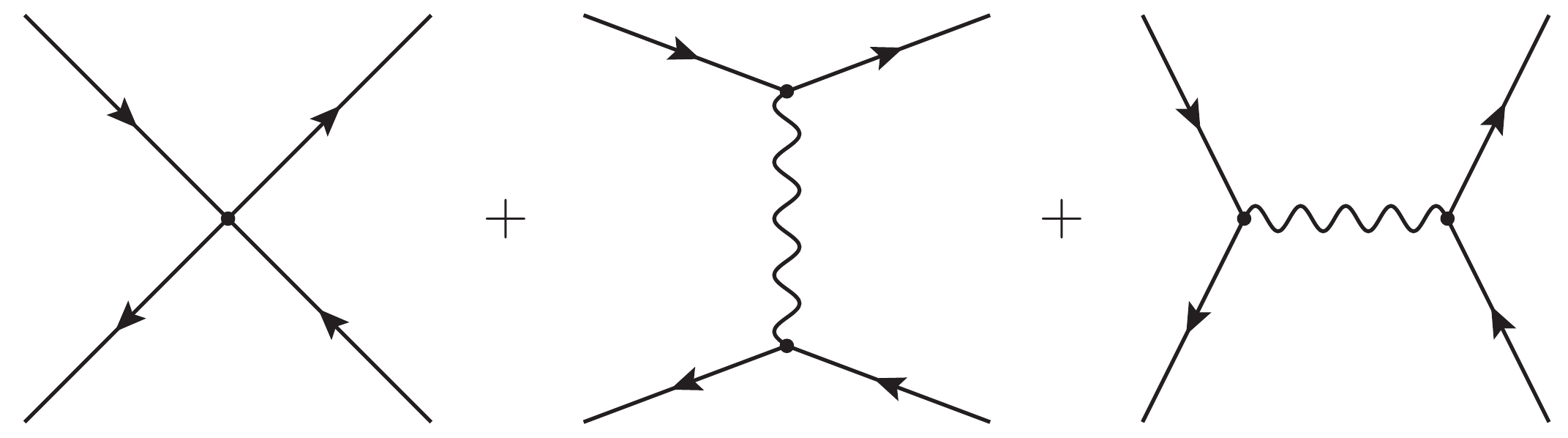}  &
		\begin{center}$\Big(4\lambda+e^{2}\dfrac{s^2+t^2+u^{2}}{st}\Big)^{2}$\end{center}
		\\ 
		% 3 row
		\begin{center}3\end{center}& \begin{center} Compton scattering $\phi \gamma \longrightarrow  \phi \gamma $ \end{center}&
		\includegraphics[height=1.4cm]{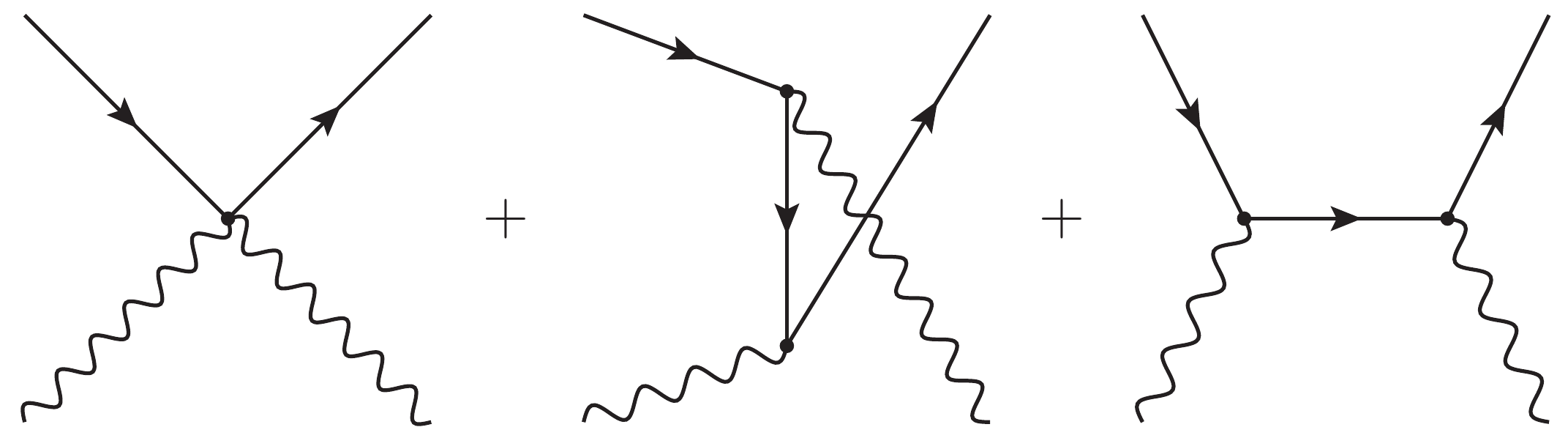}  &
		\begin{center}$4e^{4}$\end{center}
		\\ 
		% 4 row
		\begin{center}4\end{center}& \begin{center}2-photon annihilation $\phi \bar{\phi} \longrightarrow  \gamma\gamma $ \end{center}&
		\includegraphics[height=1.4cm]{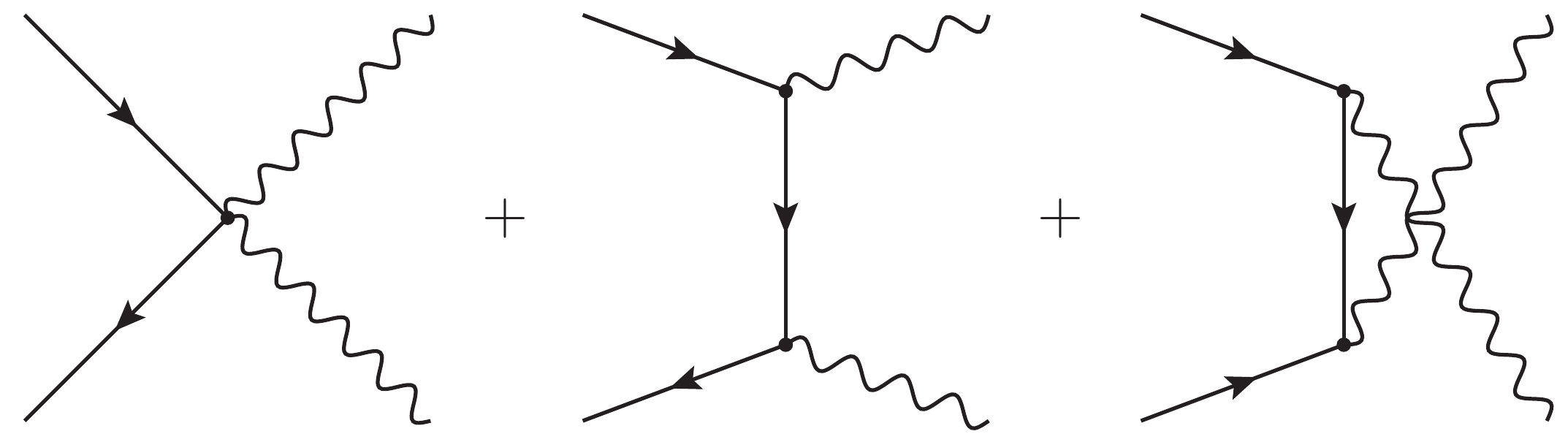}  &
		\begin{center}$8e^{4}$\end{center}
		\\ \hline\hline
	\end{tabular}
\caption{\label{tab-sqed} Binary tree-level processes in scalar QED.}
\end{table}

Inspecting the last column in Table~\ref{tab-sqed}, one could notice that the matrix elements for the Compton scattering and pair annihilation take constant values. This originates from accidental cancellation of all momentum dependent terms in the limit of massless scalar particles. If we take into account their finite masses, the matrix element for the Compton scattering would take the form:
\begin{equation}
\overline{|\mathcal{M}|^{2}}=4e^{4}\left[1+2m^{2}\left(\frac{1}{s-m^{2}}+\frac{1}{t-m^{2}}\right)+2m^{4}\left(\frac{1}{s-m^{2}}+\frac{1}{t-m^{2}}\right)^{2}\right].
\end{equation}
In the IR region, we have $t= -q_{\perp}^{2}+O(q^3/k,qm^2/k)$, where $\mathbf{q}_{\perp}=\mathbf{q}-\mathbf{v}(\mathbf{v}\cdot\mathbf{q})$ is the momentum transfer projection perpendicular to the velocity $\mathbf{v}$ of the incoming scalar particle. Thus, taking the massless limit, we neglect the terms of the form $m^{2}/(q_{\perp}^{2}+m^{2})$ which do not lead to IR singularities.

As it was shown in Refs.~\cite{Baym:1997,Arnold:2000}, the leading-log result comes from the singular behavior of the scattering matrix elements at small momentum transfer. From this we conclude that the processes of Compton scattering and pair annihilation will not contribute to the leading-log result. We would like to emphasize that the absence of IR singularities in the matrix elements of these processes is a distinctive feature of scalar QED. In contrast to this, in the case of spinor electrodynamics (see Sec.~\ref{sec-general} below or Refs.~\cite{Ahonen:1996,Ahonen:1998,Arnold:2000}), the corresponding matrix elements contain divergences of the form $\propto e^{4} s/t$, and give a contribution to the collision integral in the leading logarithmic order.

Now, let us pay attention to the scattering processes which have a singular behavior for small momentum transfer. They are shown in the first two rows in Table~\ref{tab-sqed}. The M{\o}ller scattering involves identical particles, therefore, we have to include the additional factor of $1/2$ to avoid double counting in the final state. On the other hand, its matrix element contains a singularity in the small momentum transfer region in the $t$-channel as well as in the $u$-channel. Swapping the outgoing momenta $\mathbf{k}'\leftrightarrow \mathbf{p}'$ in the $u$-channel diagram we obtain the same result as in the $t$-channel. Thus, we can neglect both the symmetry factor $1/2$ and the additional contribution from the $u$-channel and take into account only the small-$t$ expression. Using the identity $s+t+u=0$ valid in the massless case, we take only the most singular small-$t$ contributions to the matrix elements, which appear to have the same form:
\begin{equation}
\label{matrix-elem-t-channel}
|\mathcal{M}^{\phi,b}_{\phi,b}(KP\to K'P')|^{2}\simeq 4e^{4}\frac{s^{2}}{t^{2}},\quad t\to 0,
\end{equation}
where $b$ denotes $\phi$ for the M{\o}ller scattering or $\bar{\phi}$ for the Bhabha process.

Therefore, we can write the kinetic equation suitable for obtaining the leading-log result for the transport coefficients in scalar QED. Denoting $W^{\phi}=W$ and $W^{\bar{\phi}}=\bar{W}$, we have:
\begin{multline}
\label{Boltzmann-sQED}
v\cdot\partial_{X}W(X,\mathbf{k})+\mathbf{v}\cdot\mathbf{F}=-\int\frac{d^{3}\mathbf{k}'}{(2\pi)^{3}}\frac{d^{3}\mathbf{p}}{(2\pi)^{3}}\frac{d^{3}\mathbf{p}'}{(2\pi)^{3}}\frac{4e^{4} s^{2}/t^{2}}{16\epsilon_{k}\epsilon_{k'}\epsilon_{p}\epsilon_{p'}}\\
\times(2\pi)^{4}\delta^{(4)}(K+P-K'-P') n_{B}(\epsilon_{p})(1+n_{B}(\epsilon_{p'}))\frac{1+ n_{B}(\epsilon_{k'})}{1+ n_{B}(\epsilon_{k})}\\
\times\left[2W(X,\mathbf{k})-2W(X,\mathbf{k}')+W(X,\mathbf{p})-W(X,\mathbf{p}')+\bar{W}(X,\mathbf{p})-\bar{W}(X,\mathbf{p}')\right].
\end{multline}

Let us simplify Eq.~(\ref{Boltzmann-sQED}) considering the case of spatially homogeneous electric field $\mathbf{E}$ applied to the system which implies $\bar{W}=-W$. In order to extract the divergent part of the collision integral it is convenient to introduce the momentum transfer $Q=K-K'=(\omega,\,\mathbf{q})$ with the notation $\omega=\epsilon_{k}-\epsilon_{k-q}$, and express the integrand in terms of this quantity:
\begin{multline}
e(\mathbf{v}\cdot\mathbf{E})=-4\pi e^{4} \int\frac{d^{3}\mathbf{p}}{(2\pi)^{3}}\int\frac{d^{3}\mathbf{q}}{(2\pi)^{3}} \frac{\epsilon_{k}\epsilon_{p}(1-\mathbf{v}\cdot\mathbf{v'})^{2}}{\epsilon_{k-q}\epsilon_{p+q}(\omega^{2}-q^{2})^{2}}\delta(\epsilon_{k}+\epsilon_{p}-\epsilon_{p+q}-\epsilon_{k-q})\\
\times  n_{B}(\epsilon_{p})(1+n_{B}(\epsilon_{p+q}))\frac{1+n_{B}(\epsilon_{k-q})}{1+n_{B}(\epsilon_{k})}\left[W(\mathbf{k})-W(\mathbf{k}-\mathbf{q})\right], \label{kinetic-eq-W-simpl}
\end{multline}
where $\mathbf{v}'=\mathbf{p}/|\mathbf{p}|$.

The matrix element (\ref{matrix-elem-t-channel}) has a usual Rutherford $1/q^4$ behavior which is singular at small momentum transfer. However, the collision integral also contains the difference of the distribution functions with close momenta. As we will see further, this helps to reduce the severe powerlike divergence of the $q$-integral to a more weak logarithmic one. This residual divergence is usually cut by introducing the so-called \textit{Coulomb logarithm} which comes from the soft integration cutoff on $eT$ scale (for more details about the Coulomb logarithm in different plasma systems, see classical textbooks \cite{Landau-v10,Chen-book}). We will discuss the origin of this cutoff below, but now let us emphasize that the leading-log behavior of the collision integral comes from the region of small momentum transfer $q$ and in order to obtain the result in the leading-log approximation we have to expand the integrand at small $q$ keeping only the leading and first subleading terms.

One more remark concerning the small-$q$ expansion should be added in the case where the scattering particles are bosons, like in scalar QED. This expansion is based on the hierarchy $eT \lesssim q \ll p, k\sim T$ and involves the terms of the form $q/p$, $q/k$. This does not cause any problems if the resulting integrals over $p$ and $k$ are convergent. However, in the bosonic case, the equilibrium distribution functions are singular at small momenta. One has to be very accurate in this region because of the two overlapping singularities: the small-$q$ expansion becomes incorrect in the region of small $p$. Nevertheless, there is a safe way to overcome these difficulties. The prescription is to use the symmetrized variables, e.g. $p+q/2$ and $p-q/2$ or the elliptic coordinates defined in Appendix~\ref{sec-app-details} instead of usual $p$ and $p-q$. This would give the expansion which, in the subleading order, does not contain the IR-divergent $1/p$ terms (they appear only in $q^2$ correction which we are not interested in). The details of calculation can be found in Appendix~\ref{sec-app-details}. After all, we obtain the kinetic equation in the following form:
\begin{equation}
e(\mathbf{v}\cdot\mathbf{E})=-\alpha^{2}T^{3}\int\frac{dq}{q^{3}}\int d\Omega_{q} \left(1-\frac{3q}{2\pi^{2}T}+\frac{\mathbf{q}\cdot\mathbf{v}}{2T}[2n_{B}(\epsilon_{k})+1]+O(q^{2})\right)\left[W(\mathbf{k})-W(\mathbf{k}-\mathbf{q})\right], \label{kinetic-eq-W-simpl-3}
\end{equation}
where $\alpha=e^{2}/(4\pi)$ is the fine structure constant. This result was derived without any assumption about the kinetic function $W(\mathbf{k})$. To proceed further we need to specify at least its dependence on the direction of $\mathbf{k}$. For this purpose, we at first apply the constant flow velocity approximation and obtain a simple analytic result for the conductivity. Later, we will find the exact solution of the kinetic equation and check the validity of our approximation.

\subsection{Constant flow velocity approximation}
\label{subsec-flow-velocity}

The simplest approximation which allows one to specify the functional dependence of the kinetic function is the constant flow velocity approximation (CFV). It was previously used in Ref.~\cite{Baym:1997} to calculate the conductivity of plasma in usual spinor electrodynamics. Let us assume that the external electric field leads to a stationary flow of the plasma particles with some constant velocity $\mathbf{u}$, and write the distribution function of such a system as the equilibrium Bose-Einstein distribution boosted along the flow (we consider a sufficiently weak electric field so that the flow velocity is nonrelativistic):
\begin{equation}
f_{\rm CFV}(\mathbf{k})\approx f_{\rm eq}(\epsilon_{k}-\mathbf{u}\cdot\mathbf{k})\approx f_{\rm eq}(\epsilon_{k})-\frac{\partial f_{\rm eq}}{\partial\epsilon_{k}} \mathbf{u}\cdot\mathbf{k}.
\end{equation}
Comparing this expression with the definition of $W$-function (\ref{definition-W}), we immediately obtain the following ansatz:
\begin{equation}
\label{W-flow-velocity}
W(\mathbf{k})=-\mathbf{k}\cdot\mathbf{u}, \qquad \mathbf{u}={\rm const}.
\end{equation}
Substituting it into Eq.~(\ref{kinetic-eq-W-simpl-3}), we obtain the following:
\begin{equation}
e(\mathbf{v}\cdot\mathbf{E})=\frac{\alpha^{2}T^{2}}{2}[2n_{B}(\epsilon_{k})+1]\int_{q_{\rm min}}^{q_{\rm max}}\frac{dq}{q}\int d\Omega_{q} (\hat{\mathbf{q}}\cdot\mathbf{u})(\hat{\mathbf{q}}\cdot\mathbf{v})=\frac{2\pi\alpha^{2}T^{2}\ln\Lambda}{3}[2n_{B}(\epsilon_{k})+1](\mathbf{v}\cdot\mathbf{u}), \label{kinetic-eq-W-flow-velocity}
\end{equation}
where we introduced the Coulomb logarithm $\ln\Lambda=\ln(q_{\rm max}/q_{\rm min})$. As we previously discussed, the severe powerlike IR divergence of the $q$-integral is partially ameliorated and transformed into weaker logarithmic one. However, both divergences, at the upper integration limit as well as at the lower one, are the  consequences of the simplifications made by us during the calculation. We can estimate the value of the Coulomb logarithm by determining the scales on which our approximations are not valid. 

First of all, we should mention that the divergence at the upper limit is simply due to the fact that we neglected $q$ in the arguments of the Bose-Einstein distribution functions assuming it much less than the typical hard momentum, $q\ll T$. This allowed us to perform the integration over $p$ but simultaneously made the dependence of the integrand on $q$ exponentially unsuppressed at large values. This suppression would be restored if we perform the integration exactly. Thus, we can estimate the upper integration limit as $q_{\rm max}\sim T$.

The situation at the lower limit is more subtle. If we use the free-particle propagators in the matrix elements, this divergence is unavoidable because the photon has zero mass and the range of the electromagnetic interaction is infinite. However, this is only true for the particles scattering in vacuum. In the plasma, the situation changes drastically due to the screening by the thermal bath of the particles. Indeed, in order to describe the scattering process in the plasma, we should take into account the modification of the particle propagator due to the thermal effects. It is well known that in the thermal bath the particles change their dispersion relations and become quasiparticles with finite lifetime. All major features can be captured in the HTL approximation \cite{Weldon:1982,Braaten:1990,Thoma:2009}. In particular, we are interested in the thermally renormalized photon propagator because it mediates the M{\o}ller and Bhabha scatterings giving the leading contribution to the collision integral. Then, the matrix element would take the form: 
\begin{equation}
\label{matrix-elem-exact}
|\mathcal{M}(KP\to K'P')|^{2}=16e^{4}\epsilon_{k}^{2}\epsilon_{p}^{2}\left|\Delta^{HTL}_{L}(Q)+\left(\mathbf{v}\cdot\mathbf{v}'-\frac{(\mathbf{q}\cdot\mathbf{v})(\mathbf{q}\cdot\mathbf{v}')}{\mathbf{q}^{2}}\right)\Delta^{HTL}_{T}(Q)\right|^{2},
\end{equation}
where the longitudinal and transverse components of the propagator are the following
\begin{equation}
\Delta^{HTL}_{L}(Q)=\frac{1}{q^{2}-\Pi_{00}}, \qquad \Delta^{HTL}_{T}(Q)=\frac{1}{q_{0}^{2}-q^{2}-\Pi_{T}}
\end{equation}
and the components of the polarization operator equal
\begin{equation}
\Pi_{00}(Q)=-m_{D}^{2}+m_{D}^{2}\frac{q^{0}}{q}\mathcal{Q}_{0}(q^{0}/q), \quad \Pi_{T}(Q)=\frac{1}{2}\left(m_{D}^{2}-\frac{Q^{2}}{q^{2}}\Pi_{00}(Q)\right).
\end{equation}
Here $\mathcal{Q}_{0}(x)=\frac{1}{2}\ln\left|\frac{1+x}{1-x}\right|-i\frac{\pi }{2}\Theta(1-x^{2})$ is the Legendre $Q$-function of the zeroth order and $m_{D}=eT/\sqrt{3}$ is the Debye mass. It is easy to see that $\Pi_{00}$ contains the constant term $-m_{D}^{2}$ which gives the mass to the longitudinal photon (the Debye screening), while the transverse component does not have the massive term. However, the imaginary part which is present for $|q^{0}|<q$ (and corresponds to the Landau damping) also does not allow the denominator to vanish \cite{Baym:1991}. In both cases, the propagator does not contain the IR divergence $\propto 1/q^2$ any more because it is regularized on the $eT$ scale. It can be regarded as the lower cutoff for the $q$-integration leading to the following value of the Coulomb logarithm:
\begin{equation}
\ln\Lambda\approx \ln\frac{T}{m_{D}}\approx \frac{1}{2}\ln\alpha^{-1}.
\end{equation}

A weak logarithmic divergence of the collision integral in Eq.~(\ref{kinetic-eq-W-flow-velocity}) allows us to obtain the leading order result by simple estimate of the integration limits. However, such a power counting would be insufficient if the divergence was powerlike. In this situation we would be able to estimate only the parametric dependence on the coupling constant and the prefactor would remain completely unknown. This may happen if a cancellation between the kinetic functions in the collision integral does not occur. Such a situation takes place, for example, in the calculation of the color conductivity in QCD where the kinetic functions carry also color indices. This leads to more severe divergences which can be treated only by considering the exact matrix element (\ref{matrix-elem-exact}) and results in $\sim\alpha_{s} \log\alpha_{s}^{-1}$ behavior of the collision integral, where $\alpha_{s}=g_{s}^{2}/(4\pi)$ is the strong interaction coupling constant \cite{Blaizot:1999,Blaizot:2002,Bodeker:1998,Arnold:1999a,Arnold:1999b}. In this case, not only the power of coupling constant is different but also the logarithm originates from completely different range of scales, namely the interval $g_{s}^{2}T<q<g_{s}T$ gives the leading contribution to this result. 

Now, let us return to the calculation of the conductivity. In order to extract the flow velocity from Eq.~(\ref{kinetic-eq-W-flow-velocity}), we multiply it by $\mathbf{k}\, n_{B}(\epsilon_{k})(1+n_{B}(\epsilon_{k}))$ and integrate over $\mathbf{k}$. This gives us the following result:
\begin{equation}
\label{velocity}
\mathbf{u}=\frac{18\zeta(3)}{\pi^{3}\alpha^{2}\ln\alpha^{-1}T^{2}}e\mathbf{E}.
\end{equation}

Substituting the kinetic function (\ref{W-flow-velocity}) into Eq.~(\ref{electric-current}), we obtain the following simple analytic expression for the conductivity:
\begin{equation}
\label{conductivity-flow-velocity}
\sigma_{\rm CFV}=\frac{3^{2}2^{4}\zeta^{2}(3)}{\pi^{4}}\frac{T}{\alpha\ln\alpha^{-1}}\approx 2.1361\frac{T}{\alpha\ln\alpha^{-1}}.
\end{equation}
In the next subsection we will find the exact solution for the kinetic equation (\ref{kinetic-eq-W-simpl-3}) and check how close is this simple analytic result to the correct value of the conductivity.

Let us compare our result (\ref{conductivity-flow-velocity}) in the case of scalar charge carriers with the conductivity of plasma with one type of charged fermions. In Ref.~\cite{Baym:1997} it was calculated in the CFV approximation, however, the authors did not take into account the Compton scattering and pair annihilation processes which are equally important as the M{\o}ller and Bhabha processes. Their result $\sigma=(3\zeta(3)/\ln 2) \ T/(\alpha\ln\alpha^{-1})\approx 5.203\ T/(\alpha\ln\alpha^{-1})$ is more than two times larger compared to ours. The missing scattering channels were properly taken into account in Ref.~\cite{Arnold:2000}, and the corresponding CFV value for the conductivity equals $\sigma\approx 2.496\ T/(\alpha\ln\alpha^{-1})$, which is $\approx 17\%$ larger than the CFV result (\ref{conductivity-flow-velocity}) for the scalar case.

Before going further, we would like to point out that in the case of fermionic charge carriers considered in Ref.~\cite{Baym:1997}, in order to extract the flow velocity, the authors multiplied both parts of the kinetic equation by $\mathbf{v} n_{B}(\epsilon_{k})(1+n_{B}(\epsilon_{k}))$, i.e. one power of $k$ less than we did. In our case we cannot do the same, because the resulting integral over $\mathbf{k}$ would contain the logarithmic IR divergence caused by Bose-Einstein distribution functions. If regularized on the soft scale, this divergence would lead to different parametric dependence of the conductivity on the coupling constant, namely, $\sigma\propto T/(\alpha\ln^2\alpha^{-1})$. In the next subsection we will see that this is not the case and this divergence is an artifact of the method. In order to avoid it, we multiplied the kinetic equation by an additional power of $k$ before integration.

\subsection{Exact solution} 
\label{subsec-exact}

Now let us return to Eq.~(\ref{kinetic-eq-W-simpl-3}) and find its exact solution. First of all, we note that the kinetic function $W$ must be linear in electric field $\mathbf{E}$ in order to satisfy the equation. The only scalar combination which satisfies this requirement is $\mathbf{k}\cdot\mathbf{E}$. Residual dependence on $\mathbf{k}$ can only be isotropic, because there is no other vector quantities in the system. Therefore, we will look for the solution in the following form:
\begin{equation}
\label{W-ansatz}
W(\mathbf{k})=-e(\mathbf{k}\cdot\mathbf{E})g(\epsilon_{k}),
\end{equation}
where $g(\epsilon_{k})$ is the only unknown scalar function. Ansatz (\ref{W-ansatz}) is just the same as considered in Ref.~\cite{Arnold:2000} where the conductivity in the fermionic case was computed. The CFV approximation considered in the previous subsection is a partial case of Eq.~(\ref{W-ansatz}) with $g(\epsilon_{k})={\rm const}$. 

We now substitute ansatz (\ref{W-ansatz}) into Eq.~(\ref{kinetic-eq-W-simpl-3}) keeping the leading term, linear in $q$, as well as the subleading one:
\begin{multline}
(\mathbf{v}\cdot\mathbf{E})=\alpha^{2}T^{3}\int\frac{dq}{q^{3}}\int d\Omega_{q} \left(1-\frac{3q}{2\pi^{2}T}+\frac{\mathbf{q}\cdot\mathbf{v}}{2T}[2n_{B}(\epsilon_{k})+1]+O(q^{2})\right)\\
\times\left[(\mathbf{q}\cdot\mathbf{E}) g+(\mathbf{k}\cdot\mathbf{E})(\mathbf{q}\cdot\mathbf{v})g'-(\mathbf{k}\cdot\mathbf{E})\frac{q_{\perp}^{2}}{2k}g'-(\mathbf{q}\cdot\mathbf{E})(\mathbf{q}\cdot\mathbf{v})g'-(\mathbf{k}\cdot\mathbf{E})(\mathbf{q}\cdot\mathbf{v})^{2}\frac{g''}{2}+O(q^{3})\right]. \label{kinetic-eq-W-exact-1}
\end{multline}
It is easy to see that in the product of two brackets the linear order in $q$ all vanishes  after integration over the solid angle and the nonzero result comes from the terms of order $q^{2}$. Therefore, just like in the previous subsection, the integral over the absolute value of momentum transfer $q$ contains logarithmic divergence which can be cut at certain energy scales leading to the Coulomb logarithm. All higher orders in $q$ should not be taken into account since we are interested only in the leading-log behavior of the collision term. Finally, after the integration, we obtain the following equation:
\begin{equation}
1=-\frac{\pi\alpha^{2}\ln\alpha^{-1}T^{3}}{3}\left[\epsilon_{k}g''(\epsilon_{k})+4g'(\epsilon_{k})-\frac{2n_{B}(\epsilon_{k})+1}{T}(\epsilon_{k}g'(\epsilon_{k})+g(\epsilon_{k}))\right].
\end{equation}
This is an ordinary differential equation for the unknown function $g(\epsilon_{k})$. Unfortunately, it cannot be solved analytically. In order to find its numerical solution, it is convenient to introduce the new dimensionless function $\chi(x)$ defined as follows:
\begin{equation}
\label{definition-chi}
g(\epsilon_{k})=\frac{3}{\pi\alpha^{2}\ln\alpha^{-1}T^{2}}\chi(x), \qquad x=\frac{\epsilon_{k}}{T},
\end{equation}
where the dimensionful prefactor was chosen so that to obtain the resulting equation in the simplest possible form:
\begin{equation}
\label{eq-chi}
\chi''(x)+\chi'(x)\left(\frac{4}{x}-{\rm coth}\frac{x}{2}\right)-\chi(x)\frac{1}{x}{\rm coth\,}\frac{x}{2}=-\frac{1}{x}.
\end{equation}

Substituting the kinetic function (\ref{W-ansatz}) with $g$-function in the form given by Eq.~(\ref{definition-chi}) into the expression for the electric current (\ref{electric-current}), we can express the conductivity in terms of the function $\chi$ as follows:
\begin{equation}
\label{conductivity-chi}
\sigma=\frac{T}{\alpha\ln\alpha^{-1}}\frac{4}{\pi^{2}}\int_{0}^{\infty}\frac{x^{3}e^{x}}{(e^{x}-1)^{2}}\chi(x)dx.
\end{equation}

Before solving Eq.~(\ref{eq-chi}) numerically, it is useful to investigate the asymptotic behavior of its solution in the two limiting cases, $x\to 0$ and $x\to\infty$. At $x\to 0$ the differential equation takes the form:
\begin{equation}
\chi''(x)+\frac{2}{x}\chi'(x)-\frac{2}{x^{2}}\chi(x)=-\frac{1}{x},
\end{equation}
whose general solution reads as
\begin{equation}
\chi(x)=\frac{1}{3}x\ln\frac{1}{x}+C_{1}x+\frac{C_{2}}{x^{2}}.
\end{equation}
The last term should be excluded by setting $C_{2}=0$ because it would lead to a divergence in conductivity (\ref{conductivity-chi}). The second term is much smaller than the first one in the limit $x\to 0$ and can be neglected. Therefore, the asymptotical behavior of the solution at small momenta is given by
\begin{equation}
\label{asym-small}
\chi(x)\sim \chi_{0}(x)=\frac{1}{3}x\ln\frac{1}{x}, \quad x\to 0.
\end{equation}

In the opposite case of large arguments, we have the following equation, up to exponentially suppressed terms:
\begin{equation}
\label{eq-chi-large}
\chi''(x)+\chi'(x)\left(\frac{4}{x}-1\right)-\frac{1}{x}\chi(x)=-\frac{1}{x}.
\end{equation}
Its general solution is the following:
\begin{equation}
\chi(x)=1+C_{1}\left(\frac{1}{x}+\frac{2}{x^{2}}+\frac{2}{x^{3}}\right)+C_{2}\frac{e^{x}}{x^{3}}.
\end{equation}
Again, the last term is not allowed because it gives the divergent integral in conductivity (\ref{conductivity-chi}), and the second term is $o(1)$ at $x\to\infty$ and thus is negligible compared to the first term. Therefore, we have the asymptotic at large values of the argument:
\begin{equation}
\label{asym-large}
\chi(x)\sim \chi_{\infty}(x)=1, \qquad x\to\infty.
\end{equation}

Finally, we have the differential equation (\ref{eq-chi}) with the boundary conditions (\ref{asym-small}) and (\ref{asym-large}) which completely determine the unique solution. We compute it numerically and show in Fig.~\ref{fig-numerical} by the blue solid line together with approximate solution $\chi_{\rm CFV}(x)=6\zeta(3)/\pi^{2}$ extracted from the CFV approximation (\ref{velocity}) by the red dashed line.

We now substitute the solution $\chi(x)$ into Eq.~(\ref{conductivity-chi}) and find the numerical value of the conductivity:
\begin{equation}
\label{conductivity-exact}
\sigma=2.3825\, \frac{T}{\alpha\ln\alpha^{-1}}.
\end{equation}

It is interesting to compare this value with the leading-log result for the conductivity of plasma with one type of spin-$1/2$ charge carriers, calculated in Ref.\cite{Arnold:2000}. It is worth noting that in fermionic case one has to take into account also the Compton scattering and pair annihilation processes in the collision integral (for the details, see Sec.~\ref{sec-general}). Even though the additional scattering channels are present, the final result equals $\sigma\approx 2.498\, T/(\alpha\ln\alpha^{-1})$ which is still $\approx 5\%$ larger compared to our result (\ref{conductivity-exact}) for scalar particles. 

Exact integration of the differential equation gives the kinetic function only in the numerical form. However, for some purposes it would be useful to obtain a simple approximate analytical expression for it. This can be done using the variational method described in the next subsection.

\subsection{Variational calculation}
\label{subsec-variational}

Equation (\ref{eq-chi}) can be regarded as the Euler-Lagrange equation which determines the extremum of the following functional:
\begin{equation}
\label{functional-scalar}
Q[\chi]=\int_{0}^{\infty}dx\,\frac{x^{2}e^{x}}{(e^{x}-1)^{2}}\left[x\chi(x)-\frac{1}{2}(x\chi'+\chi)^{2}-\chi^{2}(x)\right].
\end{equation}
This fact can be exploited in order to calculate the conductivity using variational approach. Decomposing the unknown function $\chi$ over a finite set of physically reasonable trial functions, we can transform the problem of numerical integration of Eq.~(\ref{eq-chi}) into the problem of solving a linear algebraic system on the expansion coefficients. Indeed, if we write
\begin{equation}
\label{chi-decomp}
\chi(x)=\sum_{n=1}^{N}p_{n}y_{n}(x),
\end{equation}
then, the functional $Q[\chi]$ becomes a quadratic function of the unknown parameters $p_{n}$:
\begin{equation}
Q(\vec{p})=\sum_{n=1}^{N}B_{n}p_{n}-\frac{1}{2}\sum_{m,n=1}^{N}A_{mn}p_{m}p_{n},
\end{equation}
where the coefficients have the form:
\begin{equation}
B_{n}=\int_{0}^{\infty}\frac{x^{3}e^{x}\,dx}{(e^{x}-1)^{2}}y_{n}(x),\quad A_{mn}=\int_{0}^{\infty}\frac{x^{2}e^{x}\,dx}{(e^{x}-1)^{2}}\left[(xy'_{m}+y_{m})(xy'_{n}+y_{n})+2y_{m}(x)y_{n}(x)\right].
\end{equation}

The extremal (maximal) value of the function $Q$ is achieved for the solution of the following linear system
\begin{equation}
\hat{A}\vec{p}=\vec{B}
\end{equation}
and it is equal
\begin{equation}
Q_{\rm max}=\frac{1}{2}\vec{B}^{T}\hat{A}^{-1}\vec{B}.
\end{equation}

Substituting decomposition (\ref{chi-decomp}) into expression (\ref{conductivity-chi}) for the conductivity, we obtain:
\begin{equation}
\sigma=\frac{T}{\alpha\ln\alpha^{-1}}\frac{4}{\pi^{2}}\vec{B}^{T}\vec{p}_{\rm max}=\frac{T}{\alpha\ln\alpha^{-1}}\frac{8}{\pi^{2}}Q_{\rm max}.
\end{equation}

It is easy to check that the 1-term ansatz with $y_{1}=1$ gives result (\ref{conductivity-flow-velocity}) in the CFV approximation. However, this simple result differs from the exact value (\ref{conductivity-exact}) by $\approx 10\%$ meaning that this is a rather crude estimate. It is interesting to note that in the fermionic case the 1-term ansatz gives much better accuracy, of order $0.1\%$ \cite{Arnold:2000}. This can be explained by the fact that the exact solution of the Boltzmann equation (either for scalars or for fermions) significantly deviates from the constant in the region of small momenta. Whereas the fermionic distribution function is not sensitive to the IR region and the contribution from small $x$ is suppressed by the powers of $x$ coming from the integration measure, the Bose-Einstein distribution is singular for $x\to 0$ and it makes the IR behavior of the solution also quite important. For larger values of the argument, the exact solution tends to a constant which is easily captured by the CFV approximation. Since we consider only the leading logarithmic contribution, the sub-log terms may also give the corrections to the result of order $O(1/\ln\alpha^{-1})\sim 10\%$ \cite{Arnold:2003}. Therefore, the accuracy of the CFV results could be regarded as sufficient in the leading-log approximation.

In principle, in order to improve the accuracy, one can choose the variational ansatz with more than one term. It is natural to choose the set of basis functions which have different behavior in the small momentum region and tend to a constant for large values of the argument. We consider the following set:
\begin{equation}
\label{var-set}
y_{n}(x)=\frac{2x^{n-1}}{1+x^{n-1}}, \quad 1\leq n\leq N.
\end{equation}

\begin{figure}[h!]
	\centering
	\includegraphics[width=0.65\textwidth]{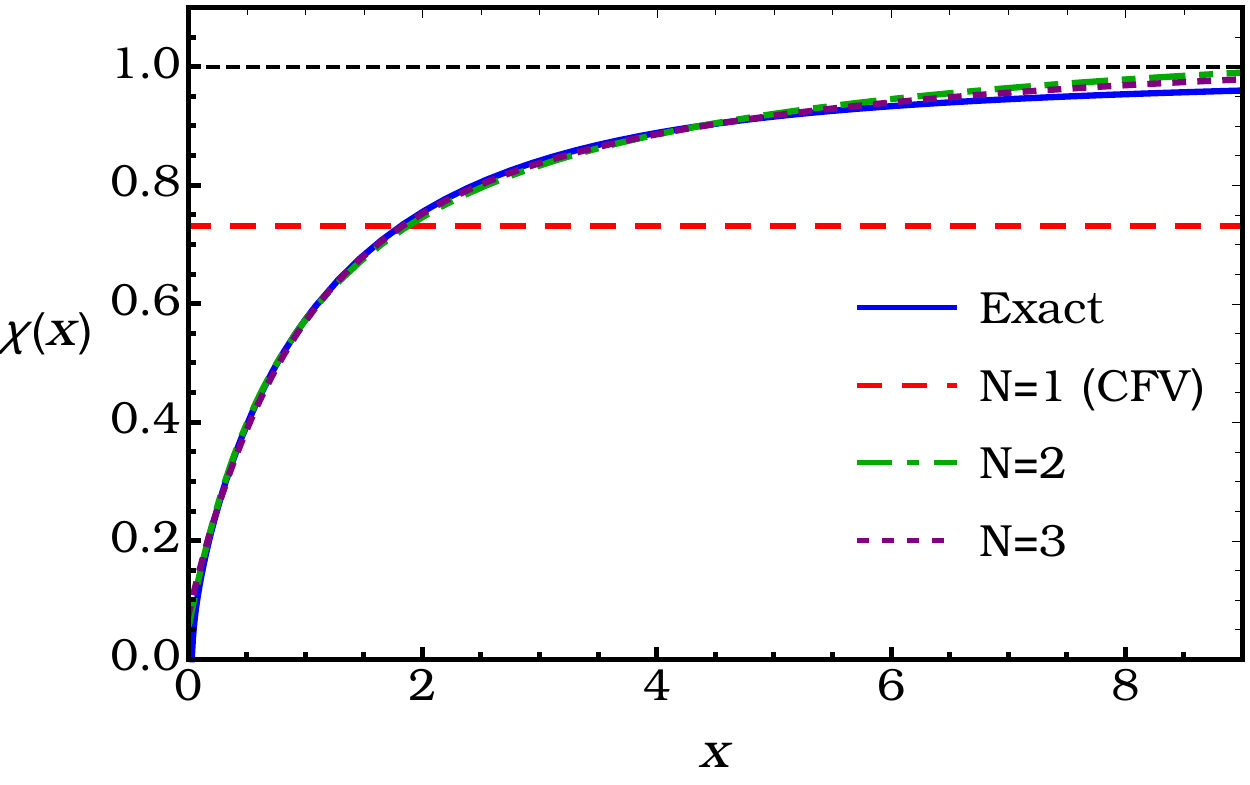}
	\caption{Exact numerical solution of Eq.~(\ref{eq-chi}) with boundary conditions (\ref{asym-small}) and (\ref{asym-large}) (blue solid line) and approximate variational solutions with different number of terms in ansatz (\ref{chi-decomp}) with trial functions (\ref{var-set}): $N=1$, i.e. the CFV approximation (red dashed line), $N=2$ (green dash-dotted line), and $N=3$ (purple dotted line). \label{fig-numerical}}
\end{figure}

We apply the variational method with a small number of terms and the results for the conductivity as well as their accuracy are listed in Table~\ref{tab-conductivity}. We would like to point out that the satisfactory accuracy is achieved already for the ansatz with two terms and this accuracy is much better than that the leading-log approximation could provide itself. We also plot the approximate kinetic functions compared to the exact solution in Fig.~\ref{fig-numerical}.

\begin{table}[!h]
	\centering
	\begin{tabular}{ c c c  c}
		\hline\hline
		$N$ & $ \{ p_{1},\ldots, p_{N}\}$ & $\vphantom{\dfrac{N}{N}} \sigma \times \frac{\alpha\ln\alpha^{-1}}{T}$ & Error (\%) \\ 
		\hline
		1 & $\vphantom{\dfrac{N}{N}}\frac{6\zeta(3)}{\pi^{2}}\approx 0.731$ & $2.1361$ & $10.34$ \\  
		2 & $\vphantom{\dfrac{N}{N}}\{ 0.049,\,0.523 \}$ & $2.3820$ & $0.02$\\  
		3 & $\vphantom{\dfrac{N}{N}}\{0.072,\,0.435,\,0.062\}$ & $2.3823$ & $0.01$ \\
		exact & $\vphantom{\dfrac{N}{N}} ...$ & $2.3825$ & $...$ \\ \hline\hline
	\end{tabular}
\caption{Electrical conductivity in scalar QED calculated by means of the variational method with ansatz (\ref{chi-decomp}) and trial functions (\ref{var-set}) with different number of terms $N$ and the error of each approximation compared to the exact value (\ref{conductivity-exact}) shown in the last row.
\label{tab-conductivity}}
\end{table}

It is interesting to note that the variational ansatz (\ref{chi-decomp}) for the distribution function with a small number of terms gives a perfect agreement with the exact kinetic function in the region of small and intermediate values of the argument although it slightly deviates from it for large $x$, see Eq.~(\ref{fig-numerical}). This can be understood as a result of exponential suppression of the integrand of functional (\ref{functional-scalar}). Visible difference between the trial function and the exact result gives exponentially suppressed contribution to the functional and to the conductivity as well.

\section{Electrical conductivity of the multicomponent Abelian plasma}
\label{sec-general}

\subsection{Degrees of freedom}
\label{subsection-dof}

Let us now consider a more general situation when a plasma contains ultrarelativistic scalar and spin-$1/2$ fermion charge carriers coupled to the $U(1)$ gauge field. Each constituent (not treating the antiparticles separately from particles) has its own electric charge $q_{a}$ and may also take part in other types of interactions. Let us assume that there are certain types of particles for which electromagnetic interaction is the most intensive, i.e. the coupling constants of all other interactions they take part are much smaller than that of the electromagnetic one. We will call these types of particles \textit{active} because they contribute to the electric current. On the other hand, if a particle can interact via more intensive interaction than the electromagnetic one, we call it \textit{passive} since its role in the conductivity is reduced to being a target for scattering of active particles. Indeed, the deviation from equilibrium for any type of particles decays in time with the characteristic relaxation time which is determined by the most intensive interaction of the particle. Therefore, the distribution of passive particles on electromagnetic scattering timescales may be regarded as an equilibrium one.

For example, in the Standard Model plasma at the temperature $T\sim 5\, {\rm GeV}$ (i.e. well below the electroweak crossover) contains 3 types of relativistic charged leptons and 4 types of quarks, of which only the leptons are active degrees of freedom and quarks are passive since they interact also strongly with $\alpha_{s}\gg\alpha$. If we consider the hypercharge conductivity in the Standard Model well above the electroweak crossover, we conclude that only right-handed leptons are active. Indeed, they interact only via the interchange of $U(1)_{Y}$ gauge bosons while other types of particles take part also in $SU(2)_{I}$ weak and $SU(3)_{c}$ strong interactions with larger couplings. From this we conclude that only fermionic degrees of freedom are important for the conductivity in the Standard Model and scalar Higgs boson never contributes to it \cite{Baym:1997,Arnold:2000,Arnold:2003}. Nevertheless, considering the plasma with both bosonic and fermionic active charge carriers might be useful for possible extensions of the Standard Model. 

Before considering the active fermions, we would like to say a few words concerning the influence of additional degrees of freedom on the purely scalar conductivity. In the region of small momentum transfer, the scattering matrix element, averaged over the spins of incoming particles and summed over the spins of outcoming particles, does not depend on the type of scatterers but only on their charge:
\begin{equation}
\label{matr-elem-small-t}
\overline{\left|\mathcal{M}^{ab}_{ab}\right|^{2}}\simeq 4e^{4}q_{a}^{2}q_{b}^{2}\frac{s^{2}}{t^{2}},\quad t\to 0.
\end{equation}

Scattering between scalars of different types contains IR divergence only in $t$-channel because the $u$-channel process does not take place for the particles of different types. On the other hand, there is no $1/2$ factor coming from taking into account of the identical particles in the final state. Therefore, scattering on the scalar particles of different type gives the same contribution as the scattering of the identical particles weighted with the squared charge of the scatterers. 

Scattering on fermions has to be considered separately. Although the matrix element has the same structure, we have to include the number of spin degrees of freedom of the target particle which gives the factor of $2$. Following the same procedure as in Sec.~\ref{subsec-collision} and Appendix~\ref{sec-app-details}, we should replace the equilibrium distributions of the target particles with the Fermi-Dirac ones. In the final result this would give a factor $1/2$ compared to the scalar case because of the integral by the Fermi-Dirac distribution functions [compare Eqs.~(\ref{int-bose}) and (\ref{int-fermi})]. Thus, we conclude that scattering on the fermions gives the same contribution as the scattering on scalars, weighted with the squared charge of scatterers. 

Finally, the kinetic equation describing the departure from equilibrium of the scalar particle distribution takes the form:
\begin{multline}
eq_{s}(\mathbf{v}\cdot\mathbf{E})=-\alpha^{2}T^{3}q_{s}^{2}N_{\rm eff}\int\frac{dq}{q^{3}}\int d\Omega_{q} \left\{1+\left(-\frac{3q}{2\pi^{2}T}\right)+\frac{\mathbf{q}\cdot\mathbf{v}}{2T}[2n_{B}(\epsilon_{k})+1]\right\}\\
\times\left[W^{s}(\mathbf{k})-W^{s}(\mathbf{k}-\mathbf{q})\right], \label{kinetic-eq-W-scalar-full}
\end{multline}
where $N_{\rm eff}=\sum_{b}q^{2}_{b}$ is the effective number of particle species weighted with the squares of their charges which is taken over all types of particles, active and passive, since they all take part in the electromagnetic scattering. The term in round brackets is present only in the sum over scalar particles. However, it has such an angular dependence that it never contributes to the conductivity in the leading-log order and can be safely omitted.

\subsection{Kinetic equation for fermions}
\label{subsec-fermions}

Let us now consider the Boltzmann equation for fermions. Table~\ref{tab-qed} shows the Feynman diagrams and matrix elements of 4 scattering processes with participating fermions of one type. 

\begin{table}[!h]
	\begin{tabular}{m{.5cm} m{4cm} m{3.3cm} m{5cm}}
		\hline\hline
		\begin{center}\textnumero \end{center} & \begin{center}Process \end{center} & \begin{center} Diagrams \end{center}
		& \begin{center}Matrix element $\overline{|\mathcal{M}|^{2}}$\end{center} \\
		\hline
		% 1 row
		\begin{center}1\end{center}& \begin{center}M{\o}ller scattering $e^{-}e^{-} \longrightarrow e^{-}e^{-}$ \end{center} &
		\vspace*{0.2cm}\includegraphics[height=1.4cm]{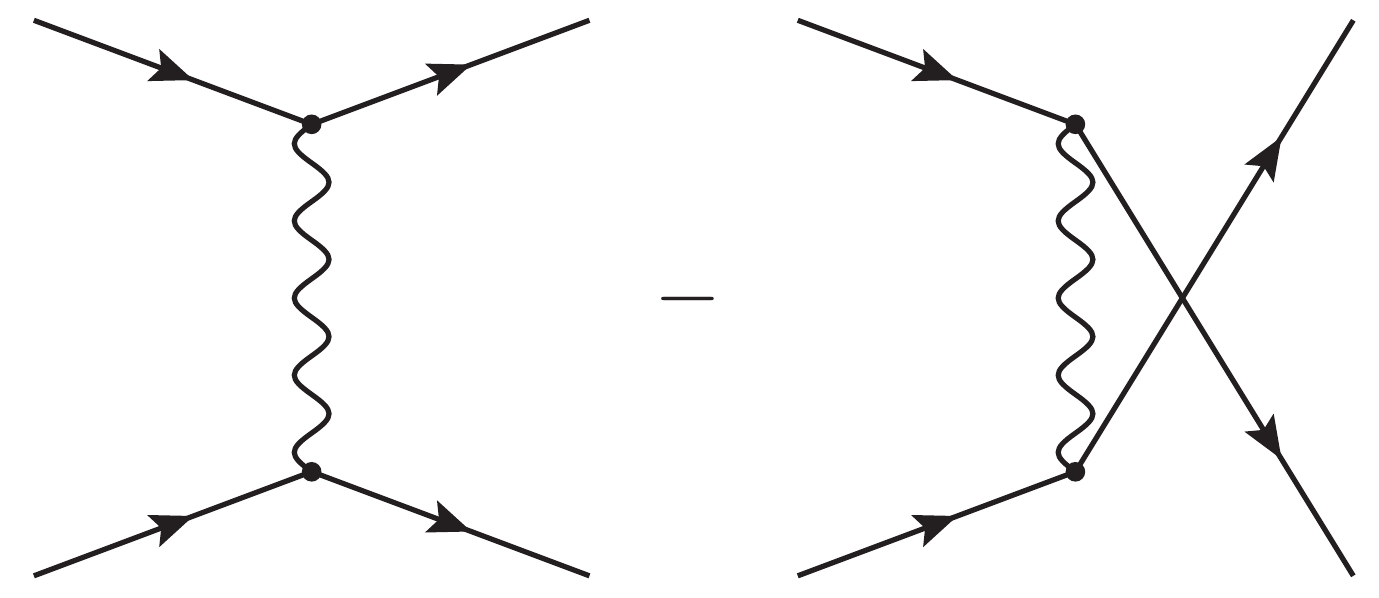}&
		\begin{center}$2e^{4} \Big(\dfrac{s^2+u^{2}}{t^{2}}+\dfrac{s^2+t^2}{u^{2}}+\dfrac{2s^2}{tu}\Big)$\end{center}
		\\ 
		% 2 row
		\begin{center}2\end{center}& \begin{center} Bhabha scattering $e^{-} e^{+} \longrightarrow  e^{-} e^{+} $ \end{center}&
		\includegraphics[height=1.4cm]{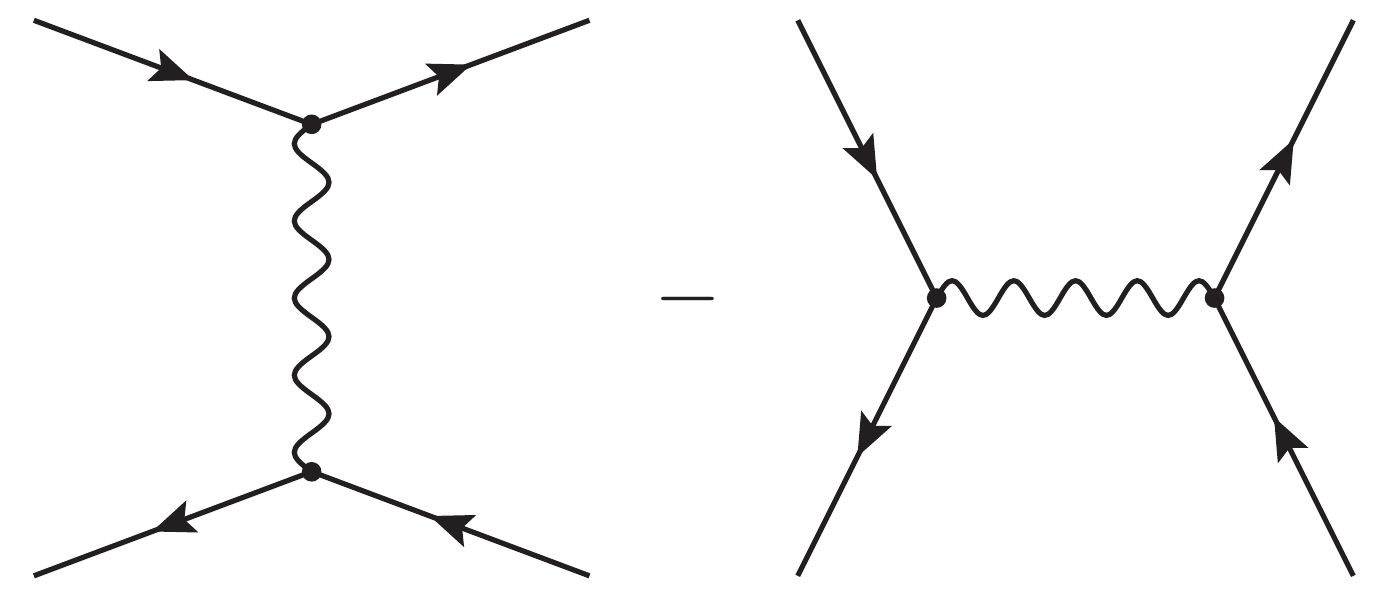}&
		\begin{center}$2e^{4} \Big(\dfrac{s^2+u^{2}}{t^{2}}+\dfrac{u^2+t^2}{s^{2}}+\dfrac{2u^2}{st}\Big)$\end{center}
		\\ 
		% 3 row
		\begin{center}3\end{center}& \begin{center} Compton scattering $e^{-} \gamma \longrightarrow  e^{-} \gamma $ \end{center}&
		\includegraphics[height=1.4cm]{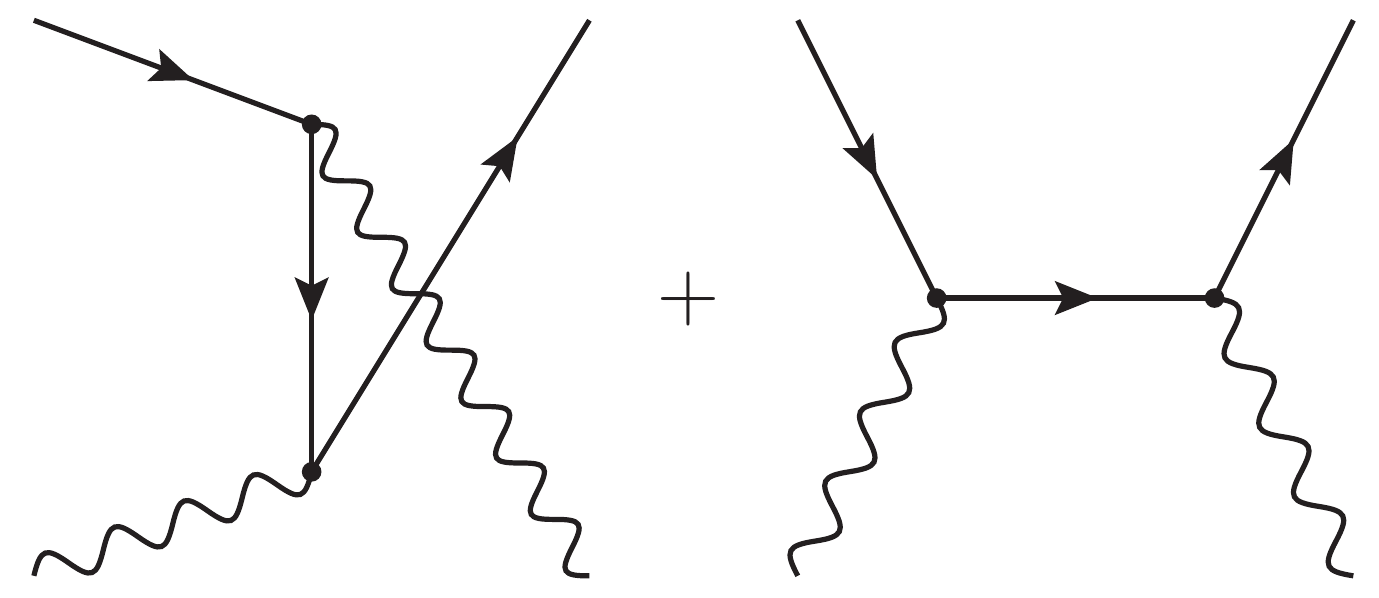} &
		\begin{center}$2e^{4}\Big(\dfrac{-u}{s}+\dfrac{s}{-u}\Big)$\end{center}
		\\ 
		% 4 row
		\begin{center}4\end{center}& \begin{center}2-photon annihilation $e^{-} e^{+} \longrightarrow  \gamma\gamma $ \end{center}&
		\includegraphics[height=1.4cm]{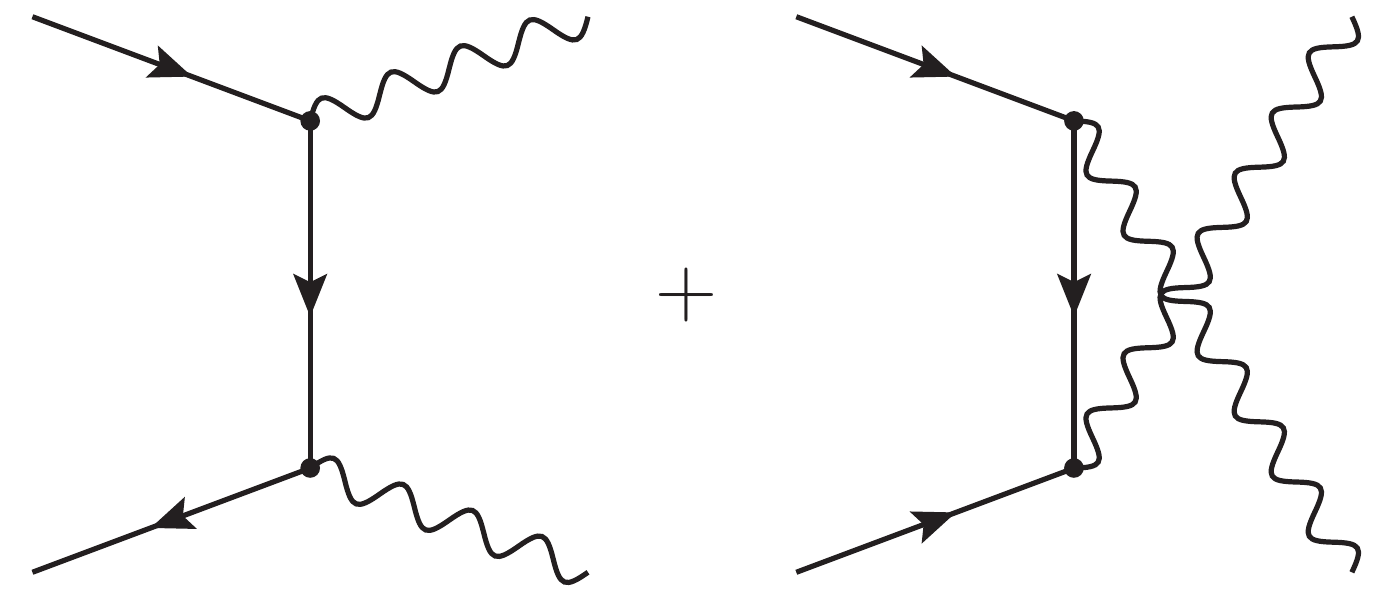} &
		\begin{center}$2e^{4}\Big(\dfrac{t}{u}+\dfrac{u}{t}\Big)$\end{center}
		\\ \hline\hline
	\end{tabular}
\caption{\label{tab-qed} Binary processes in spinor QED.}
\end{table}

Similarly to the scattering of scalar particles, the M{\o}ller matrix element contains divergences in the $t$- and $u$-channels. The latter can be led to the same form as the former by the change of variables $K'\leftrightarrow P'$. However, we must also include the factor $1/2$ connected to the fact that we have two identical particles in the final state. Therefore, we can neglect both, the additional divergence in the $u$-channel and the symmetry factor $1/2$, and take only the $t\to 0$ expression for the matrix element.

The scattering of the M{\o}ller and Bhabha types are also possible on other particle species. In this case, only the $t$-channel diagrams in the first and second rows in Table~\ref{tab-qed} are present and the matrix elements are given by the first terms in parentheses. In any case, in the region of small momentum transfer, the scattering matrix elements have the same structure (\ref{matr-elem-small-t}) for all types of particles and depend only on their charges. Therefore, we can easily derive the collision integral coming from these scattering processes following the same steps as we did in Sec.~\ref{subsec-collision} and using relations of Appendix~\ref{sec-app-details}.

In contrast to the scalar case, fermions have singular matrix elements of the Compton scattering and pair annihilation, see the third and fourth rows in Table~\ref{tab-qed}. Although this divergence is weaker, only $\propto 1/t$ for $t\to 0$, it is strong enough to contribute to the collision term in the leading-log approximation. Indeed, in the M{\o}ller scattering we have $1/t^{2}$ divergence, which is then reduced by additional powers of the momentum transfer $q$ coming from the cancellation between the kinetic functions with close momenta. However, in the Compton and pair annihilation processes only the different types of particles appear in one vertex and the above mentioned cancellation between the kinetic functions does not take place. It is also worth emphasizing that the Compton scattering and annihilation processes involve only charged particles of one type. Therefore, for a given fermion type, their contribution to the collision integral does not depend on the particle content of the model. 

Matrix element of the annihilation process $f^{+}f^{-}\to 2\gamma$  is divergent for $t\to 0$ and $u\to 0$ and the final state contains two identical particles. As before, we will take into account only $t\to 0$ singularity and omit the factor $1/2$ from the reduced phase space. Finally, the matrix element for the Compton scattering and pair annihilation have the same behavior:
\begin{equation}
\overline{\left|\mathcal{M}^{f\gamma}_{f\gamma}\right|^{2}}\simeq 2e^{4}q_{f}^{4}\frac{s}{(-t)},\quad t\to 0.
\end{equation}

Using the linearized collision integral (\ref{collision-term-linearized}), we obtain the following contribution from $f\gamma$ scattering processes:
\begin{multline}
\mathcal{C}_{f\gamma}[W^{f}]=-\frac{2\pi}{T}\int\frac{d^{3}\mathbf{p}}{(2\pi)^{3}}\frac{d^{3}\mathbf{q}}{(2\pi)^{3}}\frac{2_{s}\times 2e^{4}q_{f}^{4}  s/(-t)}{16\epsilon_{k}\epsilon_{k'}\epsilon_{p}\epsilon_{p'}} \delta(\epsilon_{k}+\epsilon_{p}-\epsilon_{k'}-\epsilon_{p'})\\
\times \left\{
n_{F}(\epsilon_{k})n_{B}(\epsilon_{p})(1+n_{B}(\epsilon_{k'}))(1-n_{F}(\epsilon_{p'}))[W^{f}(\mathbf{k})-W^{f}(\mathbf{p}')]+\right.\\
+\left.n_{F}(\epsilon_{k})n_{F}(\epsilon_{p})(1+n_{B}(\epsilon_{k'}))(1+n_{B}(\epsilon_{p'}))[W^{f}(\mathbf{k})-W^{f}(\mathbf{p})] \right\}.
\end{multline}

Since we do not have any cancellation between the kinetic functions in this case and we are interested only in the leading-log result, there is no subtlety in expanding the integrand for small momentum transfer. Keeping $q$ only in the denominator of the matrix element and neglecting it in all other factors, we can easily perform the integration. Finally, the equation describing the fermionic kinetic function takes the form:
\begin{multline}
eq_{f}(\mathbf{v}\cdot\mathbf{E})=-\alpha^{2}T^{3}q_{f}^{2}N_{\rm eff}\int\frac{dq}{q^{3}}\int d\Omega_{q} \left\{1+\frac{\mathbf{q}\cdot\mathbf{v}}{2T}[1-2n_{F}(\epsilon_{k})]\right\}\left[W^{f}(\mathbf{k})-W^{f}(\mathbf{k}-\mathbf{q})\right]-\\
-\frac{\pi}{4}q_{f}^{4}\alpha^{2}\ln\alpha^{-1}T^{2}\frac{1
+n_{B}(\epsilon_{k})}{1-n_{F}(\epsilon_{k})}\frac{W^{f}(\mathbf{k})}{k},
\label{kinetic-equation-fermions}
\end{multline}
where the first term on the right-hand side originates from the scattering involving only charged particles [an analog of the collision integral for scalars in Eq.~(\ref{kinetic-eq-W-scalar-full})] while the second one comes from the Compton scattering and pair annihilation processes.

\subsection{Result for the conductivity}
\label{subsec-conductivity-general}

Now, we have the general kinetic equations for scalars (\ref{kinetic-eq-W-scalar-full}) and fermions (\ref{kinetic-equation-fermions}). Making the substitution
\begin{equation}
\label{substitution}
W^{s,f}(\mathbf{k})=-eq_{s,f}(\mathbf{k}\cdot\mathbf{E})\frac{3}{q_{s,f}^{2}N_{\rm eff}\pi\alpha^{2}\ln\alpha^{-1}T^{2}}\chi_{s,f}(x), \quad x=\epsilon_{k}/T,
\end{equation}
we can rewrite them in the simplest form. The equation for the scalar distribution function appears to be the same as in pure scalar QED case, Eq.~(\ref{eq-chi}), whereas the equation for the fermion kinetic function takes the form:
\begin{equation}
\label{eq-chi-fermion}
\chi''_{f}(x)+\chi'_{f}(x)\left(\frac{4}{x}-{\rm th}\frac{x}{2}\right)-\frac{1}{x}\left({\rm th}\frac{x}{2}+\frac{3q_{f}^{2}}{4N_{\rm eff}}{\rm coth}\frac{x}{2}\right)\chi_{f}(x)=-\frac{1}{x}.
\end{equation}

Although this equation can be solved numerically for any given $N_{\rm eff}$ and particle charge $q_{f}$, it is more convenient to reformulate this task as a variational problem. It is straightforward to check that Eq.~(\ref{eq-chi-fermion}) determines the extremum of the following functional:
\begin{equation}
\label{functional-fermionic}
Q_{f}[\chi_{f}]=\int_{0}^{\infty}dx\,\frac{x^{2}e^{x}}{(e^{x}+1)^{2}}\left[x\chi_{f}(x)-\frac{1}{2}(x\chi'_{f}+\chi_{f})^{2}-\chi^{2}_{f}(x)\left(1+\frac{3q_{f}^{2}}{4N_{\rm eff}}\frac{x}{2}{\rm coth}\frac{x}{2}\right)\right].
\end{equation}
This functional up to a general prefactor coincides with that derived in Ref.~\cite{Arnold:2000} in a different way, see Eq.~(4.3) there. The corresponding functional for the scalar particles distribution was derived previously in Eq.~(\ref{functional-scalar}).

Knowing the distribution functions for all active degrees of freedom, we can use Eq.~(\ref{electric-current}) to calculate the electric current. Extracting the coefficient in front of the electric field, we obtain the following total conductivity:
\begin{equation}
\label{conductivity-total}
\sigma_{\rm tot}=\frac{T}{\alpha\ln\alpha^{-1}}\left[\frac{4}{\pi^{2}N_{\rm eff}}\sum_{s}^{\rm active}\int_{0}^{\infty}dx\,\frac{x^{3}e^{x}}{(e^{x}-1)^{2}}\chi_{s}(x)+\frac{8}{\pi^{2}N_{\rm eff}}\sum_{f}^{\rm active}\int_{0}^{\infty}dx\,\frac{x^{3}e^{x}}{(e^{x}+1)^{2}}\chi_{f}(x)\right].
\end{equation}

Solving Eqs.~(\ref{eq-chi}) and (\ref{eq-chi-fermion}) one can use formula (\ref{conductivity-total}) to find the exact value of the conductivity in the leading-log approximation. However, we find it useful to derive a simple analytic formula for the conductivity using the CFV approximation which is equivalent to the momentum-dependent relaxation time approximation considered in Ref.~\cite{Chakraborty:2011}. Making the simplest variational ansatz $\chi_{s,f}=A_{s,f}={\rm const}$, we calculate functionals (\ref{functional-scalar}) and (\ref{functional-fermionic}):
\begin{equation}
Q_{s}(A_{s})=6\zeta(3) A_{s}-\frac{1}{2}\pi^{2}A_{s}^{2}, \quad \Rightarrow\quad A_{s}^{\rm max}=\frac{6\zeta(3)}{\pi^{2}},
\end{equation}
\begin{equation}
Q_{f}(A_{f})=\frac{9}{2}\zeta(3) A_{f}-\frac{\pi^{2}}{4}A_{f}^{2}\left(1+\frac{3q_{f}^{2}\pi^{2}}{32 N_{\rm eff}}\right), \quad \Rightarrow\quad A_{f}^{\rm max}=\frac{9\zeta(3)}{\pi^{2}}\left(1+\frac{3q_{f}^{2}\pi^{2}}{32 N_{\rm eff}}\right)^{-1}.
\end{equation}
Finally, substituting these expressions into Eq.~(\ref{conductivity-total}), we obtain the following formula:
\begin{equation}
\label{sigma-tot}
\sigma_{\rm tot}=\frac{T}{\alpha\ln\alpha^{-1}}\left[\frac{2^{4}3^{2}\zeta^{2}(3)N_{s}}{\pi^{4}N_{\rm eff}}+\frac{2^{2}3^{4}\zeta^{2}(3)}{\pi^{4}N_{\rm eff}}\sum_{f}^{\rm active}\left(1+\frac{3q_{f}^{2}\pi^{2}}{32 N_{\rm eff}}\right)^{-1}\right],
\end{equation}
where $N_{s}$ is the number of active scalar particle species. According to Eq.~(\ref{sigma-tot}), the total conductivity is a sum over all active charged particle species $\sigma_{\rm tot}=\sum_{a}\sigma_{a}$, while each $\sigma_{a}$ contains in the denominator the contributions from all scattering channels involving particle $a$. This is in accordance with previous studies \cite{Arnold:2000,Gavin:1985,Chakraborty:2011}.

\begin{table}[!h]
	\centering
	\begin{tabular}{l  c  c c}
		\hline\hline
		System & CFV & Exact LLO & Reference \\ 
		\hline
		Spinor QED ($N_{s}=0$, $N_{f}=1$) &  2.4963 & 2.4982 & \cite{Arnold:2000} \\  
		Scalar QED ($N_{s}=0$, $N_{f}=1$) &  2.1361 & 2.3825 & [this paper]\\  
		Mixed plasma ($N_{s}=N_{f}=1$)  & 2.7110  &  2.8350 & [this paper]\\ 
		SM ($N_{s}=0$, $N_{f}=3$, $N_{\rm eff}=20/3$) & 1.8992  & 1.9054 & \cite{Arnold:2000}  \\ \hline\hline
	\end{tabular}
\caption{Electrical conductivity multiplied by $(\alpha \ln\alpha^{-1})/T$ for different ultrarelativistic plasmas calculated in the CFV approximation (\ref{sigma-tot}) as well as the exact leading logarithmic order (LLO) result.
\label{tab-comparison}}
\end{table}

In order to compare conductivities of different ultrarelativistic plasmas, we collect the CFV results as well as the corresponding leading-log results in Table~\ref{tab-comparison}. In particular, setting in Eq.~(\ref{sigma-tot}) $N_{s}=N_{\rm eff}=1$ and $N_{f}=0$, we recover, naturally, the result (\ref{conductivity-flow-velocity}) of the pure scalar QED.
%On the other hand, for one type of fermions of unit charge, we obtain $\sigma_{f}\approx 2.496\, T/(\alpha\ln\alpha^{-1})$. 
In the case of mixed plasma with one scalar and one fermion particle species of unit charge, we have even larger value for the conductivity than in pure scalar or pure fermion cases. Whereas the scalar contribution to the conductivity is simply proportional to the relative content of scalar degrees of freedom in the plasma $N_{s}/N_{\rm eff}$, the fermionic contribution has a more complicated dependence on $N_{\rm eff}$. That is why the conductivity of the mixed plasma does not simply equal to the average of the conductivities of pure scalar and fermion plasmas. This would be true only in the case when all charge carriers are active, have equal charges, and in the limit of very large number of particle species $N_{\rm eff}\gg 1$, when the expression in the parentheses in Eq.~(\ref{sigma-tot}) could be replaced with unity.

It is also interesting to consider the particular case of the Standard Model (SM) plasma well below the electroweak crossover and for temperatures $T\ll \left(\frac{\alpha_{EM}}{G_{F}}\right)^{1/2} = \frac{2^{1/4}}{\pi^{1/2}}\frac{e}{g}M_{W}\sim 25\,{\rm GeV}$ at which the weak processes are suppressed with respect to the electromagnetic ones. There we have $N_{s}=0$ active charged scalars, active fermions are represented by charged leptons because for such temperatures their weak interaction is less intensive than the electromagnetic one; several flavors of quarks (depending on temperature) are passive since they interact also strongly. Taking into account that all leptons have the same charge $q_{f}=1$, we obtain the following result:
\begin{equation}
\label{conductivity-SM}
\sigma_{\rm SM}=\frac{T}{\alpha\ln\alpha^{-1}}\frac{2^{2}3^{4}\zeta^{2}(3)N_{l}}{\pi^{4}N_{\rm eff}}\left(1+\frac{3\pi^{2}}{32 N_{\rm eff}}\right)^{-1}
\end{equation}
with the effective number of charged degrees of freedom
\begin{equation}
N_{\rm eff}=N_{l}+3_{c}\times \left(\frac{2}{3}\right)^{2} N_{u}+3_{c}\times \left(-\frac{1}{3}\right)^{2}N_{d},
\end{equation}
where $N_{l}$, $N_{u}$, and $N_{d}$ are the numbers of relativistic leptons, up and down quarks, respectively. This result is shown in the last row in Table~\ref{tab-comparison} for the case of 3 leptons ($e$, $\mu$, $\tau$) and 5 quarks ($u$, $d$, $s$, $c$, $b$) which is realized in the SM plasma at temperatures $5\,{\rm GeV}\lesssim T\lesssim 25 {\rm GeV}$. Equation~(\ref{conductivity-SM}) exactly coincides with that derived in Ref.~\cite{Arnold:2000}.

\section{Conclusion}
\label{sec-conclusion}

In this work, we studied the electrical conductivity of a general hot Abelian plasma with ultrarelativistic scalar and fermionic charge carriers in the framework of kinetic theory. The leading-log contribution of charged fermions has been studied previously in Refs.~\cite{Baym:1997,Arnold:2000} in application to the conductivity in the early Universe. However, the scalar counterpart, to the best of our knowledge, has not been considered earlier.

In the case of pure scalar electrodynamics with one type of scalar charged particles, we analyzed different binary scattering processes and arrived at conclusion that only the $t$-channel M{\o}ller and Bhabha scatterings contribute to the conductivity in the leading-log approximation. The matrix elements of these processes reveal the usual Rutherford-type divergence in the limit of small momentum transfer which is a consequence of the long-range electromagnetic interaction mediated by the free photons. It is worth noting that in Abelian plasma this divergence is reduced to merely logarithmic one due to the additional cancellation between the kinetic functions with close momenta in the collision integral. In order to deal with this divergence, we took into account the fact that the photon quasiparticles in the plasma acquire the finite lifetime due to Landau damping and obtain a finite mass due to Debye screening (longitudinal components).

The weak logarithmic divergence of the collision integral allowed us not to use the exact expression for the photon propagator with HTL corrections but simply estimate the momentum scale at which the regularization due to thermal effects takes place, $p\sim eT$. This procedure is well-known in plasma physics and introduces the Coulomb logarithm $\ln\Lambda=\ln (C/e)$ in the expressions for different plasma characteristics, in particular, in the conductivity. It gives only the leading parametric dependence on the coupling constant, however, does not allow one to determine the value of the constant $C$ inside the logarithm. In order to do this, one has to include all other binary processes as well as $1\leftrightarrow 2$ contributions and to use the HTL-corrected expressions for the propagators. For fermionic plasma this was done in Ref.~\cite{Arnold:2003}. However, computing this in the scalar case lies beyond the scope of our paper.

We simplified the Boltzmann equation for scalar particles in the leading-log approximation and found its solution using several methods. In order to take a simple analytic insight into the problem, we applied the constant flow velocity approximation assuming that the external electric field induces a uniform flow of charged particles with a velocity which does not depend on particles momentum. Following the derivation of Ref.~\cite{Baym:1997}, with a modification for the bosonic case, we obtained the simple expression (\ref{conductivity-flow-velocity}) for the conductivity. Naturally, it has the same parametric dependence on the coupling constant, although the numerical prefactor has a lower value compared to the fermionic case considered in Refs.~\cite{Baym:1997,Arnold:2000}.

The axial symmetry of the problem allowed us to reduce the functional dependence of the kinetic function to one unknown scalar function of the absolute value of momentum. We derived the second order differential equation determining this function and found its numerical solution. The result is shown by the blue solid line in Fig.~\ref{fig-numerical} and it significantly deviates from the constant for small arguments. Since we deal with bosons whose distribution functions are sensitive to the IR region, this deviation makes an impact on the value of the conductivity and leads to a $\sim 10\%$ error of the CFV result compared to the exact value. 

The exact numerical integration of the kinetic equation can be also reformulated in the form of variational principle, in analogy with Ref.~\cite{Arnold:2000}. Decomposing the kinetic function over a finite set of trial functions we found a satisfactory approximation for the exact solution. We showed that even the ansatz with two terms in the decomposition gives an error of order $0.02\%$ providing much better accuracy than that of the leading-log approximation itself. 

Finally, we generalized our results to the case of a multicomponent Abelian plasma with an arbitrary number of scalar and fermionic particles coupled to the $U(1)$ gauge field. Depending on the strength of interactions of charged particles, apart from the $U(1)$ interaction, the particle species were classified into active (those which contribute to the electric current) and passive (scatterers for active species). We calculated the contribution of the passive charge carriers to the collision integral and concluded that they do not modify its general structure but only multiply the whole expression by an effective number of scatterers weighted with the squares of their charges.  Furthermore, we derived the kinetic equation for the fermionic charge carriers including in the collision integral not only the M{\o}ller and Bhabha scattering but also the Compton scattering and pair annihilation processes which have in this case the singular cross sections. In order to find a solution of the obtained Boltzmann equation, we reformulated it as a variational problem, and the corresponding functional (\ref{functional-fermionic}) is in full accordance with Ref.~\cite{Arnold:2000}. Applying the simplest variational ansatz, we obtained the most general formula for the conductivity of the multicomponent plasma (\ref{conductivity-total}) which includes all possible partial cases, e.g. the pure scalar QED considered in this article, as well as the case of the Standard Model plasma addressed in Ref.~\cite{Arnold:2000}. 

Even though, the Standard Model plasma does not contain the active ultrarelativistic scalar charge carriers, the formula derived in our paper may be useful in some extensions of the Standard Model or for the comparison with the results of lattice simulations of Abelian field theories containing scalars. The simulations conducted in Refs.~\cite{Figueroa:2018a,Figueroa:2018b,Figueroa:2019} were aimed at calculating the rate of the chirality nonconservation in the external magnetic field in Abelian gauge theory with a scalar field. The theoretical prediction for this rate can be derived in the framework of magnetohydrodynamics and the result is expressed in terms of electrical conductivity of the medium. Using our results for the conductivity in hot scalar QED, one would obtain the rate of chirality breaking which is roughly 10 times smaller than the numerical result of the simulations \cite{Figueroa:2019}. The explanation of this discrepancy may shed light on the mechanism of the chirality nonconservation in this system and the role of the scalar long-wavelength modes in this process. We plan to address this issue elsewhere. 

In this work, we considered the ultrarelativistic plasma where the masses of all charge carriers are much smaller than temperature, $m\ll T$. Nevertheless, it would be interesting to study the influence of the finite mass on the conductivity. Obviously, in the nonrelativistic case with $T\ll m$, the conductivity should be exponentially suppressed $\sim \exp(-m/T)$ due to the small number of charge carriers. However, its exact functional dependence for arbitrary $m$ and $T$ requires a separate investigation. It is worth noting that massive charged scalars exist in a hadron gas for temperatures below the deconfinement phase transition of QCD. The electrical conductivity of such a plasma was considered in Refs.~\cite{Fraile:2006,Nicola:2007,Lee:2014,Greif:2016,Ghosh:2017,Atchison:2017,Kadam:2018,Ghosh:2018}. In contrast to our results, the conductivity of the pion gas decreases with temperature and reaches its minimal value in the vicinity of the phase transition. However, their results cannot be explained only by the impact of a finite mass. A very important role is also played by the strong interaction between hadrons which determines their mean-free path.

\begin{acknowledgements}
The author is grateful to Prof.~Mikhail Shaposhnikov for the illuminating discussions and valuable comments, as well as to Adrien Florio, Daniel G. Figueroa, and Prof.~Eduard Gorbar for the careful reading of the manuscript and useful remarks. This work was supported by the ERC-AdG-2015 Grant No. 694896 and by the Swiss National Science Foundation Grant No. 200020B\_182864.
\end{acknowledgements}

\appendix

\section{Details of calculation of the collision integral}
\label{sec-app-details}

In this Appendix, we present the details of calculation of the collision integral avoiding the overlapping singularities in the integrals over $\mathbf{q}$ and $\mathbf{p}$.

\begin{figure}[h!]
	\centering
	\includegraphics[width=0.35\textwidth]{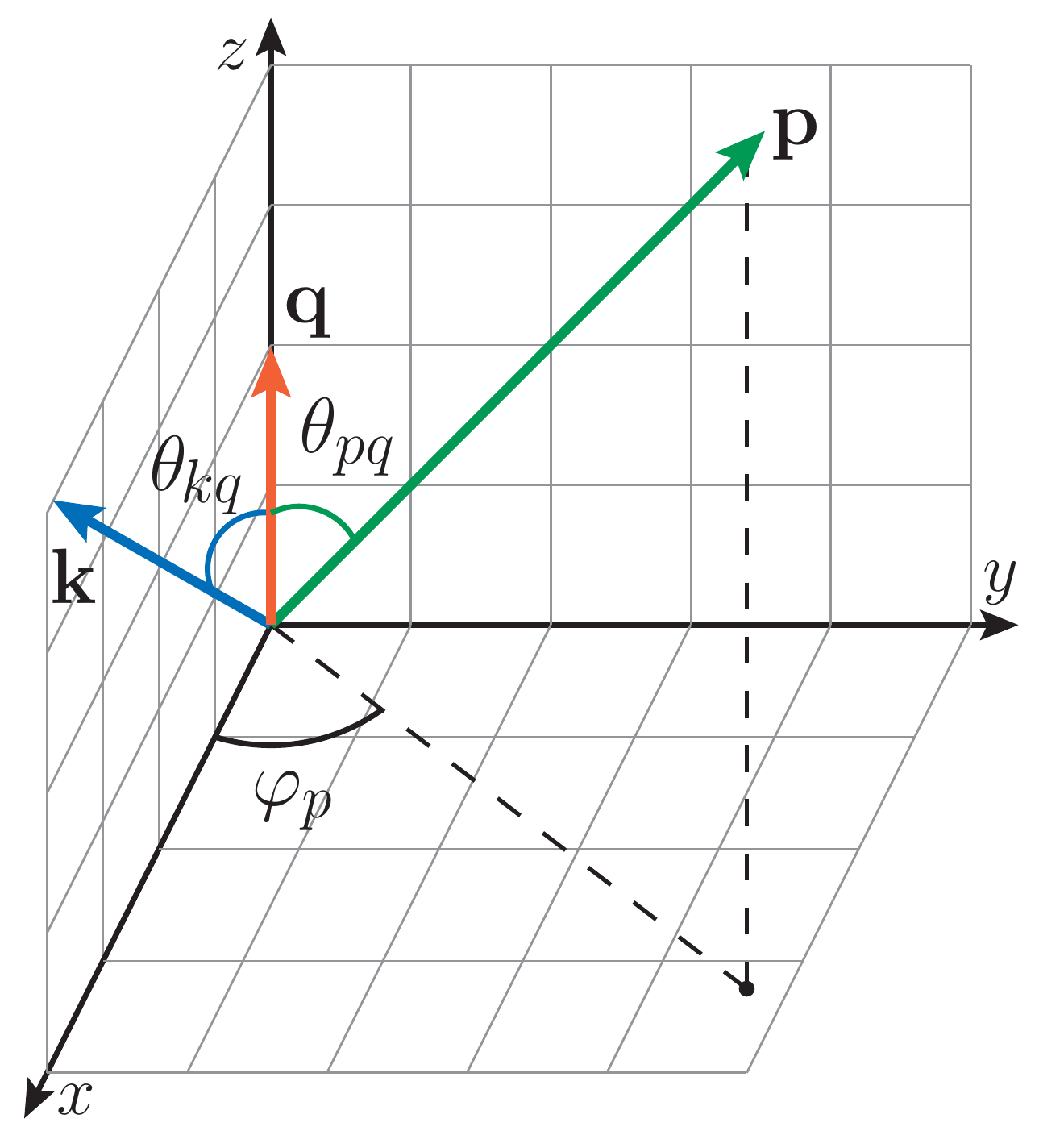}
	\caption{Geometry of the scattering. \label{fig-geometry}}
\end{figure}

We use the set of symmetric variables which are more convenient to treat the delta-function in the collision integral. Instead of three components of $\mathbf{p}$ we use the azimuthal angle $\varphi_{p}$ and two elliptic variables $\xi$ and $\eta$. The angle $\varphi_{p}$ is defined in the spherical coordinate system with $z$-axis directed along the vector $\mathbf{q}$ and the vector $\mathbf{k}$ lying in $xOz$ plane, see Fig.~\ref{fig-geometry}, and the elliptic coordinates are defined as follows:
\begin{equation}
\label{elliptic-coords}
\xi=\frac{\epsilon_{p+q}+\epsilon_{p}}{2},\qquad \eta=\frac{\epsilon_{p+q}-\epsilon_{p}}{2}.
\end{equation}
The corresponding Jacobian equals $|J|=\frac{2}{q}(\xi^{2}-\eta^{2})$. The $q$-integration is performed in a spherical coordinate system with the polar axis directed along the vector $\mathbf{k}$. The collision integral in Eq.~(\ref{kinetic-eq-W-simpl}) takes the form:
\begin{multline}
e(\mathbf{v}\cdot\mathbf{E})=-\frac{2\alpha^{2}}{\pi^{3}} \int q\,dq\int d\Omega_{q}\int_{0}^{2\pi}d\varphi_{p}\int_{q/2}^{\infty}d\xi \int_{-q/2}^{q/2}d\eta\, (\xi-\eta)^{2}\\ \times\frac{\epsilon_{k}}{\epsilon_{k-q}}\frac{1+n_{B}(\epsilon_{k-q})}{1+n_{B}(\epsilon_{k})} \frac{(1-\cos\theta_{kq}\cos\theta_{pq}-\sin\theta_{kq}\sin\theta_{pq}\cos\varphi_{p})^{2}}{[(\epsilon_{k}-\epsilon_{k-q})^{2}-q^{2}]^{2}}\\
\times \delta(2\eta-\epsilon_{k}+\epsilon_{k-q}) n_{B}(\xi-\eta)(1+n_{B}(\xi+\eta))\left[W(\mathbf{k})-W(\mathbf{k}-\mathbf{q})\right], \label{kinetic-eq-W-simpl-2}
\end{multline}

We can now easily integrate the delta-function and expand the integrand in the limit $q\to 0$ keeping only the leading and first subleading order. This implies the following expressions:
\begin{equation}
\eta=\frac{\epsilon_{k}-\epsilon_{k-q}}{2}= q\left(\frac{\cos\theta_{kq}}{2}-\frac{q\sin^{2}\theta_{kq}}{4k}+O(q^{2})\right),
\end{equation}
\begin{equation}
(\xi-\eta)^{2}=\xi^{2}\left(1-\frac{q\cos\theta_{kq}}{\xi}+O(q^{2})\right),
\end{equation}
\begin{equation}
\frac{\epsilon_{k}}{\epsilon_{k-q}}\frac{1+n_{B}(\epsilon_{k-q})}{1+n_{B}(\epsilon_{k})}= 1+q\cos\theta_{kq}\left(\frac{n_{B}(\epsilon_{k})}{T}+\frac{1}{k}\right)+O(q^{2}),
\end{equation}
\begin{equation}
\frac{\epsilon_{k}}{\epsilon_{k-q}}\frac{1-n_{F}(\epsilon_{k-q})}{1-n_{F}(\epsilon_{k})}= 1+q\cos\theta_{kq}\left(-\frac{n_{F}(\epsilon_{k})}{T}+\frac{1}{k}\right)+O(q^{2}),
\end{equation}
\begin{multline}
\int_{0}^{2\pi}d\varphi_{p} (1-\cos\theta_{kq}\cos\theta_{pq}-\sin\theta_{kq}\sin\theta_{pq}\cos\varphi_{p})^{2}\\
= 3\pi \sin^{4}\theta_{kq}\left[1+q\cos\theta_{kq}\left(\frac{1}{k}+\frac{1}{\xi}\right)+O(q^{2})\right],
\end{multline}
\begin{equation}
\frac{1}{(Q^{2})^{2}}=\frac{1}{[(\epsilon_{k}-\epsilon_{k-q})^{2}-q^{2}]^{2}}= \frac{1}{q^{4}\sin^{4}\theta_{kq}}\left(1-\frac{2q\cos\theta_{kq}}{k}+O(q^{2})\right),
\end{equation}
\begin{equation}
n_{B,F}(\xi-\eta)[1\pm n_{B,F}(\xi+\eta)]= n_{B,F}(\xi)[1\pm n_{B,F}(\xi)]\left(1+\frac{q\cos\theta_{kq}}{2T}+O(q^{2})\right).
\end{equation}

After substituting all these expressions into Eq.~(\ref{kinetic-eq-W-simpl-2}) we conclude that the $\xi$-dependence of the integrand does not contain any singular terms and the integration can be performed leading to the following result:
\begin{equation}
\label{int-bose}
\int_{q/2}^{\infty}\xi^{2}n_{B}(\xi)(1+n_{B}(\xi))d\xi=\frac{\pi^{2}T^{3}}{3}\left(1-\frac{3q}{2\pi^{2}T}+O(q^{2})\right).
\end{equation}
\begin{equation}
\label{int-fermi}
\int_{q/2}^{\infty}\xi^{2}n_{F}(\xi)(1-n_{F}(\xi))d\xi=\frac{\pi^{2}T^{3}}{6}\left[1+O(q^{3})\right].
\end{equation}
Collecting the subleading order contributions altogether, we finally obtain the kinetic equation in the form of Eq.~(\ref{kinetic-eq-W-simpl-3}).

We also list below some integrals with Bose-Einstein and Fermi-Dirac distribution functions used in the calculations of the conductivity in the main text:
\begin{equation}
\int_{0}^{+\infty}dk\, k^{p} n_{B}(k)(1+n_{B}(k))=\Gamma(p+1)\zeta(p)T^{p+1}, \quad p>1,
\end{equation}
\begin{equation}
\int_{0}^{+\infty}dk\, k^{p} n_{F}(k)(1-n_{F}(k))=\Gamma(p+1)\zeta(p)T^{p+1}\left(1-\frac{1}{2^{p-1}}\right), \quad p>-1,
\end{equation}
where $\zeta(p)$ is the Riemann zeta-function and $\Gamma(p)$ is the Euler gamma-function \cite{Gradstein:book}.

\end{document}